\documentclass[twocolumn,times]{aastex62}
\usepackage{amsmath,amstext}
\usepackage{tablefootnote}
\usepackage{comment}
\usepackage{hyperref}
\usepackage{multirow}
\hypersetup{
    colorlinks=true,
    linkcolor=blue,
    citecolor=blue,
    urlcolor=blue
}
\usepackage{bm}
\newcommand{\ergshz}{${\rm erg \ s^{-1} \ Hz^{-1}}$ }
\newcommand{\ergs}{${\rm erg \ s^{-1}}$ }
\newcommand{\ergcms}{${\rm erg \ cm^{-2} \ s^{-1}}$}
\def\uJy    {$\mu$Jy}
\def\ltsima{$\; \buildrel < \over \sim \;$}
\def\simlt{\lower.5ex\hbox{\ltsima}}
\def\gtsima{$\; \buildrel > \over \sim \;$}
\def\simgt{\lower.5ex\hbox{\gtsima}}
\newcommand{\msun}{{\rm\,M$_\odot$}}

\newcommand{\src}{{\rm\,AT2019ijn }}
\newcommand{\srcs}{{\rm\,AT2019ijn}}

\begin{document}

\title{
AT2019ijn: a fast-rising, slow-decaying blue optical transient with exceptionally bright radio emission 
}
\correspondingauthor{X. W.~Shu, \& L. M. Sun} 
\email{xwshu@ahnu.edu.cn, sunluming@ahnu.edu.cn}

\author[0009-0001-8733-2088]{Hucheng~Ding }
\affil{Department of Physics, Anhui Normal University, Wuhu, Anhui, 241002, China}

\author[0000-0002-7020-4290]{Xinwen~Shu}
\affil{Department of Physics, Anhui Normal University, Wuhu, Anhui, 241002, China}

\author[0000-0002-7223-5840]{Luming~Sun}
\affil{Department of Physics, Anhui Normal University, Wuhu, Anhui, 241002, China}

\author[0000-0002-8708-0597]{Liangduan Liu}
\affiliation{Institute of Astrophysics, Central China Normal University, Wuhan 430079, China}
\affiliation{Education Research and Application Center, National Astronomical Data Center, Wuhan 430079, China}
\affiliation{Key Laboratory of Quark and Lepton Physics (Central China Normal University), Ministry of Education, Wuhan 430079, China}

\author[0009-0002-3809-1609]{Lei Yang}
\affil{Department of Physics, Anhui Normal University, Wuhu, Anhui, 241002, China}

\author[0000-0002-1067-1911]{Yunwei Yu}
\affiliation{Institute of Astrophysics, Central China Normal University, Wuhan 430079, China}
\affiliation{Education Research and Application Center, National Astronomical Data Center, Wuhan 430079, China}
\affiliation{Key Laboratory of Quark and Lepton Physics (Central China Normal University), Ministry of Education, Wuhan 430079, China}

\author[0000-0001-5458-8354]{Xueguang Zhang}
\affiliation{Guangxi Key Laboratory for Relativistic Astrophysics, School of Physical Science and Technology, GuangXi University, Nanning, 530004, China}

\author[0009-0005-6943-7803]{Ying Gu}
\affiliation{Guangxi Key Laboratory for Relativistic Astrophysics, School of Physical Science and Technology, GuangXi University, Nanning, 530004, China}

\author[0000-0001-7171-5132]{Fangkun~Peng}
\affil{Department of Physics, Anhui Normal University, Wuhu, Anhui, 241002, China}

\author{Fabao~Zhang}
\affil{Department of Physics, Anhui Normal University, Wuhu, Anhui, 241002, China}

\author{Zhumao Zhang}
\affil{Department of Physics, Anhui Normal University, Wuhu, Anhui, 241002, China}

\author[0000-0002-2169-0472]{Ningyu Tang}
\affil{Department of Physics, Anhui Normal University, Wuhu, Anhui, 241002, China}

\begin{abstract}
We report the discovery of a peculiar optical transient, AT2019ijn, 
occurred in the nuclear region of a dwarf galaxy at $z = 0.2729\pm0.0001$.
It rises rapidly to peak at a luminosity of $M_\textit{g} = -21.05 \pm 0.02$ in $5.26\pm0.29$ days, followed by a slow decline over more than a month,  during which the optical emission has a persistently high blackbody temperature from $T_{\rm BB}= 1.45^{+0.16}_{-0.13}$ to $1.59^{+1.68}_{-0.51}\times10^4$ K.
The radio emission is exceptional which peaks at 641 days after optical discovery with a high luminosity of $2.0\pm0.1\times10^{31}$ \ergshz. 
The peak radio luminosity is at least two orders of magnitude brighter than known radio-bright fast blue optical transients and supernova explosions at similar epochs, 
but comparable to jetted tidal disruption events.
The luminous and long-lasting radio emission with a late-time peak can be explained by an off-axis relativistic jet with a 
viewing angle of $38.9^{+7.0}_{-6.1}\arcdeg$. 
We discuss possible origins for AT2019ijn and favor a jetted tidal disruption event involving an intermediate-mass black hole of $1.32^{+1.19}_{-0.67}\times10^5$\msun, although a jetted magnetar model cannot be fully ruled out.
AT2019ijn represents a new class of relativistic optical transients that highlights the importance of radio surveys for discovering off-axis jetted events.
\end{abstract}


\section{Introduction}

The development of high-cadence wide-field optical surveys in recent years, such as the Panoramic Survey Telescope and Rapid Response System \citep[Pan-STARRS,][]{Chambers2016}, Zwicky Transient Facility \citep[ZTF,][]{Bellm2019}, and Asteroid Terrestrial-impact Last Alert System \citep[ATLAS,][]{Tonry2018}, leads to the discovery of transient events with extreme parameters, for example, a new population of fast-evolving blue optical transients \citep[FBOT,][]{Drout2014}.
These transients are characterized by a rapid rise to their peak brightness within$\simlt$10 days and blue colors ($\textit{g}-\textit{r}<-0.2$) near the peak, followed by an exponential decay to the post-flare level within a month \citep{Pursiainen2018,Wiseman2020}.

Although the majority FBOTs could represent extreme cases of core-collapse supernovae \citep[CCSNe,][]{Ho2023ApJ}, a subset of luminous FBOTs (LFBOTs) with optical peak luminosities brighter than $-20$ mag  \citep{Perley2019,Perley2021,Coppejans2020,Ho2020,Yao2022} cannot be explained as a high-luminosity extension of CCSNe.
LFBOTs are extremely rare as their occurring rate is only $\lesssim0.1\%$ of CCSNe below $z\sim0.5$ \citep{Ho2023ApJ}.
They are typically hosted by dwarf star-forming galaxies with modest subsolar metallicity \citep{Klencki2025}, but most display a large offset from the galaxy's center, with a projected distance between 0.3 kpc and 5.4 kpc \citep{Chrimes2024MN}.
The physical origin of LFBOTs is still uncertain, and possible explanations for their high luminosities and short rise times include the interaction of the supernova shockwave with a dense circumstellar medium \citep[CSM,][]{Margutti2019}, the injection of energy from spin-down of a young magnetar \citep{Yu2015}, merger of binary white dwarfs \citep{Yu2019,Lyutikov2019}, tidal disruption of star by a stellar-mass black hole or neutron star binary companion \citep{Metzger2022, Tsuna2025}, or by an intermediate-mass black hole \citep[IMBH, a few $10^4$\msun,][]{Kuin2019, Perley2019,Gutierrez2024, Zhang2022RAA}. 
The latter scenario seems supported by the late-time detection of UV and possibly X-ray emission in AT2018cow, the prototype of LFBOT, which could be interpreted as the thermal emission of the slowly-evolving accretion disk around a black hole of $\sim$$10^3-10^4$\msun~ \citep{Migliori2024, Inkenhaag2025}.

\begin{figure*}[htbp!]
\epsscale{1.15}
\plottwo{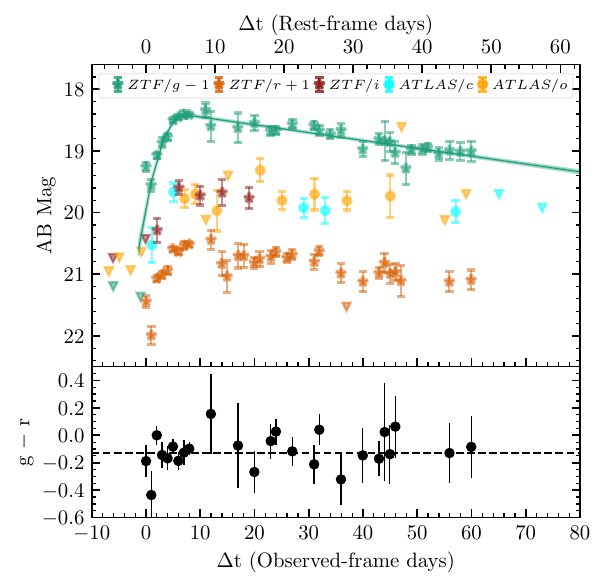}{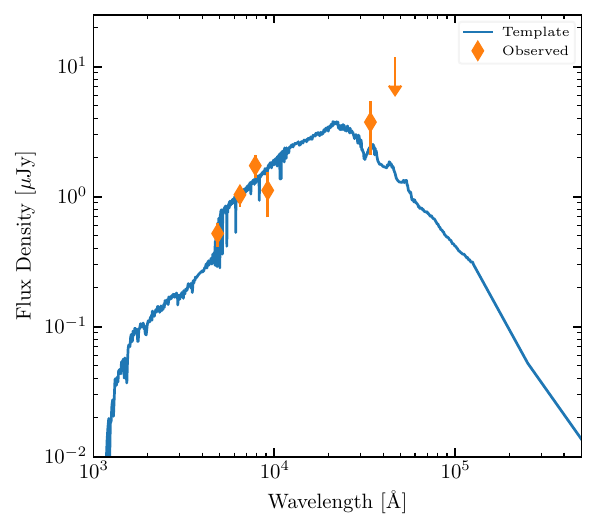}
\caption{
\label{fig:1}
Left: The host-subtracted optical light curves of AT2019ijn (upper panel). 
The phase is computed relative to the discovery date $t_0=$ 58634.24 (MJD). 
Data points with a signal-to-noise ratio (S/N) greater than 3 are considered as detections, while the rest are displayed at 3$\sigma$ upper limits as downward triangles.
We also show the best-fitting model with a Gaussian rise and an exponential decay for the g-band data. 
The $\textit{g}-\textit{r}$ color evolution is shown in the lower panel and the black dashed line represents the median $\textit{g}-\textit{r}=-$0.13.
Right: Result of SED fitting using optical to MIR photometry obtained from DESI-Legacy and WISE survey. The best-fitting galaxy template at $z = 0.273$ is indicated by the solid line.
}
\end{figure*}

While stellar tidal disruption events (TDEs) have some optical properties similar to LFBOTs, such as the high peak luminosity, blue and constant color near the peak, they evolve on much slower timescales of several months \citep{Yao2023}. 
TDEs are typically identified to be spatially coincident with the center of a galaxy. 
Such a requirement would bias against dwarf galaxies hosting IMBHs or high-$z$ TDEs, when the optical surveys  
are not deep enough. 
In addition, there is an increasing number of TDEs that could occur in off-nuclear regions \citep{Lin2018, Zhang2025NatAs, Jin2025, Li2025, Shu2025, Yao2025, Stein2026}, most of which may host IMBHs. 
More recently, \citet{Somalwar2025} reported the discovery of AT2024puz, which exhibits properties similar to both LFBOTs and TDEs. 
This suggests that previous searches of LFBOTs by typically requiring a fast rise and fade timescale of several days \citep{Ho2023ApJ} may suffer from selection effects against transients with longer evolution timescale in either rising and fading phase. 
Therefore, exploring transients with energetics and evolution timescale that are intermediate between them  
are important to link LFBOTs and TDEs, providing new insights into the nature of LFBOTs. 


In this Letter, we present the discovery and results from the multi-wavelength analysis of 
AT2019ijn, a peculiar optical transient occurred in the center of a dwarf galaxy. 
While AT2019ijn shows a fast rise-to-peak time of a few days, 
the post-peak optical emission decays slowly over more than a month, distinct from known LFBOTs. 
In addition, it displays extremely bright and long-lived radio emission, 
peaking at 641 days since the optical discovery. 
The peak radio luminosity is $2\times10^{31}$ \ergshz at 3 GHz, 
at least two orders of magnitude higher than other radio-emitting LFBOTs or supernova explosions at similar epochs, 
but comparable to jetted TDEs. 
These combined properties suggest that \src might represent a new class of optical transients ever found. 
Throughout this work, we adopt a cosmology of $\Omega_{\rm M}$ = 0.3, $\Omega_{\lambda}$ = 0.7, and $H_0$ = 70 km s$^{-1}$ Mpc$^{-1}$, and report optical magnitudes in the AB system.

\section{observation and data reduction} \label{sec:obs_data}

\subsection{Optical Flare}
\label{subsection:2.1}

We identified \src via cross-match between known optical transients from the Transient Name Server\footnote{\url{https://www.wis-tns.org}} (TNS) and bright ($>$ 5 mJy) radio transients detected in the Very Large Array Sky Survey \citep[VLASS,][]{Lacy2020}, using a match radius of 2.5$\arcsec$, which is the typical angular resolution of VLASS. 
The catalog of radio transients was constructed by performing image subtractions between different epochs of VLASS observations. 
AT2019ijn was discovered by the ZTF survey as a new optical transient on 2019 May 31 ($\rm MJD=58634.24$) with $\textit{g} = 20.26\pm0.07$ mag and then reported to TNS \citep{Nordin2019TNS}.
We adopt this MJD as the discovery date $t_{0}$ throughout this work. 
The last non-detection was reported on 2019 May 30 ($\rm MJD=58633.2$), with a 3 $\sigma$ upper limit on the 
flux of $\textit{g} > 22.37$ mag. 
The transient was also detected by ATLAS.
We retrieved the forced photometry data of ZTF \footnote{\url{https://ztfweb.ipac.caltech.edu/cgi-bin/requestForcedPhotometry.cgi/}} \citep{Masci2019} in $\textit{g}-$, $\textit{r}-$ and $\textit{i}-$ bands and ATLAS\footnote{\url{https://fallingstar-data.com/forcedphot/}} \citep{Tonry2018, Smith2020} in $\textit{c}-$ and $\textit{o}-$ bands. 
All photometry for both AT2019ijn and its host galaxy (Section \ref{subsection:2.2}) have been corrected for Milky Way dust extinction. We adopted the E(B-V) values from the dust maps of \citet{Schlafly2011} and applied extinction law of \citet{Fitzpatrick1999} with $\rm R_V = 3.1$.
According to the survey strategy of ZTF \citep{Bellm2019b}, the field of AT2019ijn was visited several times in one night with time intervals typically  less than $\sim1$ hour. We combined the photometric data of these exposures into a single data point for each night. 
The resulting light curves are shown in Figure \ref{fig:1} (left).
Note that the flux of host galaxy has been subtracted from the light curves, by using the median of baseline flux for data taken before the optical discovery ($\Delta t < -100$ days). 
Note that the first detection in the ZTF light curves displays an excess in the flux in comparison with that obtained in the subsequent observation, perhaps representing a signal from the precursor's emission. 
However, the current data quality is not sufficient to secure the excess emission, thus prevents it from further studies.

\subsection{Host Galaxy}
\label{subsection:2.2}

At the position of AT2019ijn (R.A. = $13^{h}12^{m}40\fs5512$ and decl. = +21$\arcdeg$13$\arcmin$39\farcs6520), a faint extended source is detected by the DESI Legacy Imaging Survey  \citep{Dey2019} with magnitudes of $\textit{g} = 24.61\pm0.23$, $\textit{r} = 23.86\pm0.21$, $\textit{i} = 23.30\pm 0.22$, and $\textit{z} = 23.78\pm 0.41$, respectively. 
It is likely the host galaxy of AT2019ijn because its central position coincides with the transient source, with a spatial offset by only 0.04$\arcsec$, less than the typical positioning error of $\sim0.1\arcsec$ \citep{Masci2019}.
The host galaxy was observed by Keck telescope with its Low Resolution Imaging Spectrograph \citep[LRIS,][]{Oke1995} on December 2023 (PI: Gregg Hallinan).
We reduced the data following the standard routine.
A preliminary analysis of the Keck/LRIS spectrum revealed the detections of the [O II], [O III] and H$\alpha$ emission lines at a common redshift of $z = 0.2729\pm0.0001$, measured by Gaussian fittings.
We will refer to this redshift for \src in our following analysis. 

We fitted the photometric data of the host galaxy with Fitting and Assessment of Synthetic Templates \citep[FAST;][]{Kriek2009} to infer its stellar properties.
We adopted the stellar population models of \citet{Bruzual2003} and the stellar initial mass function of \citet{Chabrier2003}, assuming delayed exponentially declining star-formation history and uniform dust extinction for all stellar populations.
In order to better constrain the model of spectral energy distribution (SED), we included the fluxes at 3.4$\mu$m and 4.6$\mu$m, which are $\textit{W}_1 = 3.75\pm1.66$ \uJy~and $\textit{W}_2 = 1.28\pm3.97$ \uJy~as measured by the Legacy Surveys forced photometry conducted on data taken with unWISE \citep{Dey2019, Meisner2017}, despite neither measurement exceeding 3$\sigma$ threshold.
The results are shown in Figure \ref{fig:1} (right).
We found that the host is a dwarf galaxy with the $\textit{g}-$band absolute magnitude of $M_{\textit{g}} = -15.8$ mag and a stellar mass of $M_{\rm gal} = 2.04_{-1.23}^{+1.27}\times10^{8}$\msun. 
The SED fitting yields a star-formation rate (SFR) of $0.16^{+1.36}_{-0.13}$\msun~yr$^{-1}$ and a corresponding specific SFR (sSFR) of $7.9^{+238}_{-6.1} \times10^{-10}$ yr$^{-1}$.
Note that the stellar mass estimated from the SED fittings is relatively robust, as the results are consistent with each other if adopting different star-formation histories or varying the metallicity across $Z=0.004-0.05$. 
In addition, we also estimated a stellar mass of $5.5\times10^{8}$\msun~ 
using the mass-to-light ratio \citep{Reyes2025}, which is consistent with a dwarf galaxy 
as derived from the SED fittings. 


\subsection{Radio Data}

\subsubsection{Detection of a Radio Transient from Archival Surveys}
\label{subsection:2.3.1}
Before the optical flare, the galaxy was not detected by the Faint Images of the Radio Sky at Twenty-Centimeters (FIRST) survey \citep{Becker1995} and the VLASS epoch I observations, with 3$\sigma$ upper limits of 0.32 and 0.41 mJy beam$^{-1}$ at 1.4 GHz and 3 GHz, respectively.
For radio images, we used the {\tt IMFIT} task within the Common Astronomy Software Applications \citep[CASA,][]{CASATeam2022} to model the radio emission with a two-dimensional elliptical Gaussian, which allows for determining the best-fit source position, integrated and peak flux density, and their associated uncertainties. 
We adopted a conservative error estimate, defined as the quadrature sum of the flux uncertainty from {\tt IMFIT} and 5\% systematic error \citep[e.g.,][]{Yang2022}, as the former may be underestimated.
A radio transient was clearly detected by the VLASS epoch II observations with a peak flux $f_{\rm peak}= 6.0 \pm 0.3$ mJy, about 412 days after the optical discovery. 
The position of the radio transient coincides with the optical one, indicating that they are physically connected.
The transient was also detected in the subsequent VLASS epoch III and IV data at a phase of $\Delta t=$1333 days and 2331 days, but the radio flux has dropped to $3.7 \pm 0.2$ mJy and $2.0 \pm 0.2$ mJy, respectively.

\begin{deluxetable}{ccccccccc}
\centering
\tablewidth{0pt}
\setlength{\tabcolsep}{0.8mm}
\tablecaption{Radio observations and flux measurements \label{tab:table1}}
\tablehead{
&\colhead{Observatory} & \colhead{Date} & \colhead{Phase} & \colhead{$\nu$} & \colhead{$F_\nu$} \\
& \colhead{} & \colhead{} & \colhead{(days)} & \colhead{(GHz)} & \colhead{(mJy/beam)}
}
\startdata
& FIRST & 1998 Oct  & $-$7547 & 1.4 & $<0.32$ \\ 
& VLASS & 2017 Sep 25 & $-$613 & 3.0 & $<0.41$ \\ 
&       & 2020 Jul 16 & 412 & 3.0 & $5.97\pm 0.34$ \\ 
&       & 2023 Jan 23 & 1333 & 3.0 & $3.67\pm0.21$ \\
&       & 2025 Oct 17 & 2331 & 3.0 & $1.95\pm0.19$ \\
& ASKAP & 2019 Apr 22 & $-$39 & 0.89 & $<1.6$ \\ 
&       & 2020 Dec 30 & 579 & 1.37 & $3.34\pm 0.24$ \\ 
&       & 2022 Jan 02 & 947 & 1.66 & $6.54\pm 0.37$ \\ 
&       & 2022 Mar 26 & 1029 & 0.89 & $2.40\pm 0.14$ \\ 
&       & 2023 Dec 28 & 1672 & 0.94 & $2.55\pm 0.25$ \\
&       & 2024 Nov 01 & 1981 & 1.37 & $2.85\pm 0.16$ \\
&  VLA  & 2021 Mar 02 & 641 & 3.0 & $8.61\pm 0.43$ \\ 
&       & 2022 Feb 22 & 998 & 3.0 & $5.98\pm 0.30$ \\ 
&       & 2024 Apr 30 & 1796 & 10.0 & $1.32\pm 0.06$ \\ 
&       & 2024 May 02 & 1798 & 1.5 & $2.64\pm 0.17$ \\ 
&       & 2024 May 02 & 1798 & 3.0 & $2.44\pm 0.13$ \\ 
&       & 2024 May 06 & 1802 & 6.0 & $1.75\pm 0.11$ \\ 
&  GMRT & 2024 Jul 27 & 1884 & 0.75 & $1.64\pm 0.19$ \\ 
&       & 2024 Jul 29 & 1886 & 1.26 & $2.18\pm 0.14$ \\ 
\enddata
\end{deluxetable}

The Rapid ASKAP Continuum Survey \citep[RACS,][]{McConnell2020} is a large-area radio survey covering the entire southern sky up to del.=$+41^\circ$. 
RACS enables the study of radio transients over various time-scales (from seconds to even years). This survey covers three bands: RACS-low (887.5/943.5 MHz), RACS-mid (1367.5 MHz), and RACS-high (1655.5 MHz), with a typical root-mean-square (rms) of $\sim$0.25 mJy beam$^{-1}$.
We retrieved all public RACS archival data through the CSIRO ASKAP Science Data Archive (CASDA)\footnote{\url{https://data.csiro.au/domain/casda}} to investigate radio properties of AT2019ijn. 
We measured the radio fluxes and corresponding uncertainties directly from the RACS science-ready images, following the same procedure as described for processing the VLASS data.
While there is no radio emission at the position of \src in the observation performed on 2019 Apr 22, a bright radio source was detected later on 2020 Dec 30, with a flux of $3.3\pm0.2$ mJy.
This confirms AT2019ijn as a bona fide radio transient, which displays continued raising in flux on a years-long timescale. Details on the ASKAP observations were shown in Table \ref{tab:table1}.

\subsubsection{VLA}
\src was also detected as a bright radio source in the S-band (3 GHz),  
as part of the target observations of the Karl G. Jansky Very Large Array (VLA) performed in 2021 Mar and 2022 Feb, respectively (project code: 20B-393; PI: Dillon Dong, and 21B-322; PI: Gregg Hallinan). 
To further study the evolution and origin of the years-long bright radio emission of \srcs, we conducted  
follow-up VLA observations in 2024 during B configuration (project code: 24A-078; PI: Xinwen Shu), 
at L-, S-, C- and X-bands, covering a broad frequency range of $\sim$1.5-10 GHz. 
For all VLA data, 3C 286 and J1327+2210 were used as flux and phase reference calibrator, respectively.
The data were calibrated with VLA Calibration Pipeline (version 2022.2.0.64) within CASA (version 6.4.1). 
We performed a manual inspection of the reduced data, flagging additional radio frequency interference (RFI), if present, and reran the pipeline.
Subsequently, we proceeded to image \src using the CASA task {\tt tclean} with Briggs weighting and ROBUST parameter of $-$0.01, ensuring that there are no obvious imaging artifacts. 
The flux densities and uncertainties were then measured using the same procedure as described in Section \ref{subsection:2.3.1}.
The radio emission is detected in all bands and appears unresolved with no extended emission detected in the residual map ($>3\sigma$).
This is confirmed by the ratios of the integrated and peak flux densities which are close to 1. 
Therefore, for consistency, only peak flux densities are used in
our following analysis. The VLA observational log and
flux density measurements are presented in Table \ref{tab:table1}.
Except for L-band, the source is bright enough at higher frequencies for 
sub-band flux measurements by splitting the 
Measurement Set (MS) data from independently calibrated spectral windows \citep[e.g.,][]{Dong2021}, 
with a spectral resolution of 128 MHz. 
This would allow for better sampling in the broad-band SED, hence characterizing the radio SED properties (Section 3.3). 
Note that our results would not be affected if using different spectral resolution for the sub-band splitting. 

\subsubsection{uGMRT} 

We triggered an upgraded Giant Metrewave Radio Telescope (uGMRT) DDT observations at band 5 (centered at $\sim$1.0$-$1.5 GHz) on 2024 Jul 27 and band 4 (centered at $\sim$0.7 GHz) on 2024 Jul 29, 
under project ddtC366 (PI: Xinwen Shu). 
The uGMRT observations were performed quasi-simultaneously as VLA in 2024 May, providing a more robust constraint on the broad-band SED of \srcs, 
as well as its evolution properties. 
3C 286 and J1330+251 were employed as flux calibrator and phase calibrator, respectively.

The uGMRT data were reduced using CASA (version 5.6.1), employing a pipeline adapted from the CAsa Pipeline-cum-Toolkit for Upgraded Giant Metrewave Radio Telescope data REduction \citep[CAPTURE; ][]{Kare2021}.
Initial calibration involved flagging known bad channels, followed by RFI removal using the flagdata task in both clip and tfcrop modes.
For imaging, we utilized tclean task with multi-scale multi-frequency synthesis \citep[MS-MFS, ][]{Rau2011} deconvolution, configured with two Taylor terms (nterms=2) and W-Projection \citep{Cornwell2008} to accurately model the wide bandwidth and the non-coplanar field of view of uGMRT.
The flux densities and uncertainties were measured following the same procedure presented in Section \ref{subsection:2.3.1}.
AT2019ijn was detected in all uGMRT observations as an unresolved source, and the flux density measurements are listed in Table \ref{tab:table1}.

\begin{figure*}[htbp!]
\epsscale{1.15}
\plotone{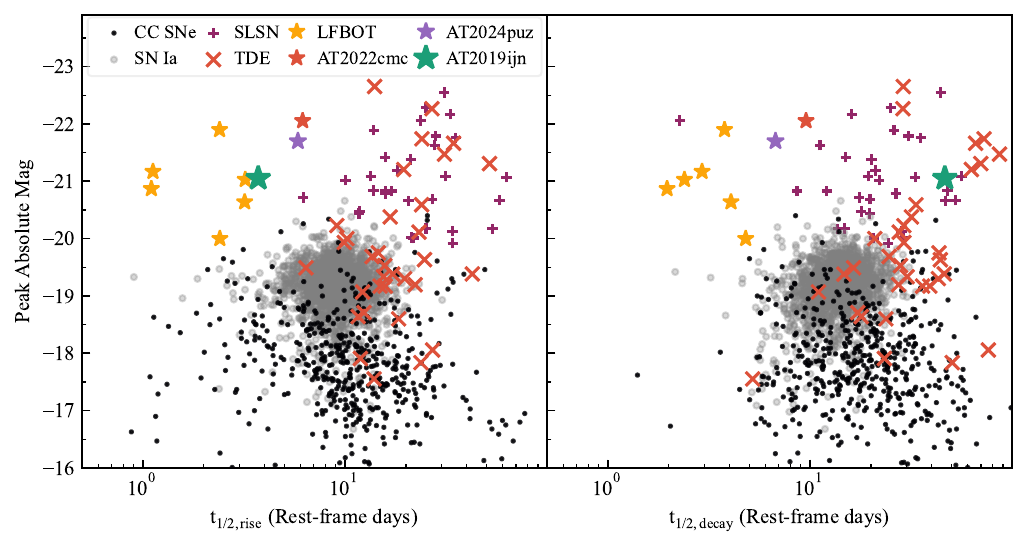}
\caption{
\label{fig:comparison}
The comparison of rest-frame optical rise time (at $\textit{g}-$band, expressed as $t_{\rm 1/2, rise}$) and the decline time ($t_{\rm 1/2, decay}$) with peak absolute magnitude of AT2019ijn to those of core-collapse SNe, SNe Ia, SLSNe from \citet{Ho2023Nature,Perley2020}, TDEs from  \citet{Yao2023}, and LFBOTs from \citet{Yao2022, Ho2023ApJ, Ho2023Nature, Pursiainen2025, Somalwar2025}. The data used for comparison are mainly obtained using filters as close to the rest-frame $\textit{g}-$band as possible. We also include the jetted TDE AT2022cmc for comparison, but for the light curve evolution of its thermal component as modeled by \citet{Hammerstein2026}. In this case, the rise time-scale is model dependent and should be treated with caution, as the early emission is dominated by a non-thermal jet emission. 
}
\end{figure*}
\subsection{Nondetections from X-ray and Gamma-Ray Observations}
\label{subsection:High Energy}
We searched for the potential GRB counterpart to AT2019ijn in \textsl{Swift} \footnote{\url{https://swift.gsfc.nasa.gov/archive/grb_table/}} and \textsl{Fermi} \footnote{\url{https://heasarc.gsfc.nasa.gov/w3browse/fermi/fermigbrst.html}} GRB catalogue. The match radius was set to the 90\% confidence error region for \textsl{Swift}/BAT and to the 68\% confidence error region for \textsl{Fermi}/GBM as listed in the respective catalogue. Although several GRB counterparts were found near the position of \srcs, a significant temporal offset ($\ge 1$ year) exists between their prompt emissions and the optical epoch of the transient \srcs. This discrepancy is inconsistent with established GRB optical emission timescales \citep{Kann2011}, leading us to reject a physical association. Following the successful identification of a GRB associated with a radio transient in the MAXI data \citep{Dong2021}, we extended our search to the MAXI GRB table \footnote{\url{https://maxi.riken.jp/grbs/}}. However, no counterpart was found.

To investigate potential X-ray flares of AT2019ijn in archival data, we utilized the High-Energy light curve Generator \footnote{\url{http://xmmuls.esac.esa.int/upperlimitserver/}} \citep{Konig2022, Saxton2022}.
The position of \src was covered by the ROSAT survey but no source was  detected, yielding a flux of $< 4\times10^{-13}$ \ergcms at 0.2 -- 12 keV.  
It was also not detected by INTEGRAL hard X-ray observations, with flux upper limits of $\lesssim4\times10^{-12}$ \ergcms at 20 -- 40 keV, 40 -- 60 keV, and 60 -- 100 keV, respectively. 
The XMM-Newton Slew Survey observed the position of AT2019ijn between 2002 and 2003, 
yielding similar upper limits on flux of $< 1.5\times10^{-12}$ \ergcms (0.2 -- 2 keV), $< 1.4\times10^{-11}$ \ergcms (2 -- 12 keV), and $< 3\times10^{-12}$ \ergcms (0.2 -- 12 keV). 
In addition, we also checked eROSITA Upper Limits Server\footnote{\url{https://erosita.mpe.mpg.de/dr1/erodat/upperlimit/single/}} \citep{Tubin2024, Merloni2024} for possible detection of \srcs. 
eROSITA has observed the position of AT2019ijn on 2019 Dec 15, $\sim$190 days after the optical peak, and did not detect the source either, providing more rigorous upper limits of $< 9.7\times10^{-14}$ \ergcms at 0.2 -- 2.3 keV and $< 1.4\times10^{-13}$ \ergcms at 0.2 -- 5 keV, respectively. 
This corresponds to an upper limit on the X-ray luminosity of $L_{\rm 0.2-5 keV}<3.3\times10^{43}$\ergs.
To summarize, no X-ray emission associated with \src was detected in archival observations. 
However, since no X-ray observations have been performed close to the optical peak, we cannot fully rule out the possibility of coincident X-ray brightening with optical one in \srcs.

\section{Analysis and Results}\label{sec:analysis}

\subsection{Optical Light Curve and Evolution}

To quantify the characteristics of the light curve evolution, we fitted the $\textit{g}-$band light curve with a model of a Gaussian rise followed by an exponential decay \citep{Yao2023} via the Markov chain Monte Carlo (MCMC) approach.
The fittings are converged by both calculating the integrated autocorrelation time ($\tau$) and visually inspecting that the walkers reaching steady states in the trace plots after burn-in, resulting in stable and well-constrained posterior distributions in the corner plot.
We employed 64 walkers, each evolving for 2000 steps (maximum autocorrelation time $\tau_{\rm max}= 45$) to sample the posterior and discarded the first 50\% of the chain as burn-in to eliminate any dependence on initial conditions.
We present the best-fitting models in Figure \ref{fig:1}, which yields a rest-frame peak phase of $\Delta t = 5.26\pm0.29$ days ($\rm MJD=58641$) since the discovery,  with the peak absolute magnitude of $M_\textit{g} = -21.05 \pm 0.02$.
According to the best-fit model, the rest-frame duration above the half-maximum in the rising phase is $t_{\rm 1/2,rise} = 3.7\pm0.2$ days, while the decay time-scale is $t_{\rm 1/2, decay} = 46.3\pm1.5$ days. 
Note that if fitting to the $\textit{r}-$band light curve, we obtained similar constraints on the evolution timescales. 
We then calculated the color by using directly the $\textit{g}$ and $\textit{r}-$band magnitudes observed in the same night, and no interpolation was applied.
Interestingly, \src exhibits a persistent blue color of $\textit{g}-\textit{r}=-0.13$ throughout the rise, peak and decline phases for over two months. 

We performed fittings to the optical SED with a blackbody emission model for epochs where detections in more than one band were available (within the same night), 
using MCMC approach with 64 walkers and 2000 steps ($\tau_{\rm max}= 1$).
The near-peak SED conforms to a blackbody with a high temperature of $T_{\rm BB}= 1.45^{+0.16}_{-0.13}\times10^4$K, corresponding to a blackbody luminosity of $L_{\rm BB} = 1.52^{+0.27}_{-0.18}\times10^{44}$\ergs. 
We will refer to the blackbody luminosity as the bolometric luminosity of \srcs. 
The blackbody temperature at a phase of $\sim47$ days post peak is still high with $T_{\rm BB}=1.59^{+1.68}_{-0.51}\times10^4$K.
The fast rise timescale, peak brightness, and persistent blue color of \src 
are reminiscent of the population of LFBOTs. 
However, the slow decay in the post-peak light curve is distinct from known LFBOTs, which typically fade over $\simlt$1 week. 
\src would likely be missed from current selection criteria for LFBOTs \citep[typically $t_{1/2}<12$ days,][]{Ho2023ApJ}, suggesting a potential selection bias toward only the most rapidly evolving transients.  



In order to gain further insights into the uniqueness of \srcs, we compared the peak absolute magnitudes, rise and decay time scales of AT2019ijn with normal SNe, SLSNe, LFBOTs and TDEs. As shown in Figure \ref{fig:comparison}, 
AT2019ijn exhibits a higher peak luminosity than normal SNe and most TDEs, while its luminosity is comparable to LFBOTs and SLSNe.
However, its rise time (a few days) is much shorter than the typical timescale of $\sim$40-day observed in most SLSNe \citep{Chen2023}. 
In addition, the blackbody temperature of $\sim$16000 K at $\Delta t$$\sim$50 days post-maximum is much higher than typical SLSNe of $<$8000 K (e.g., \citealt{Inserra2018}).
On the other hand, its decline phase is considerably longer than those of LFBOTs, but comparable to some TDEs and SLSNe. 
Despite its shared preference with both LFBOTs and SLSNe for dwarf host galaxies \citep{Klencki2025,Perley2016a,Perley2016b}, AT2019ijn appears to not fully fit comfortably into either category.  

\begin{figure*}[htbp!]
\epsscale{1.15}
\plotone{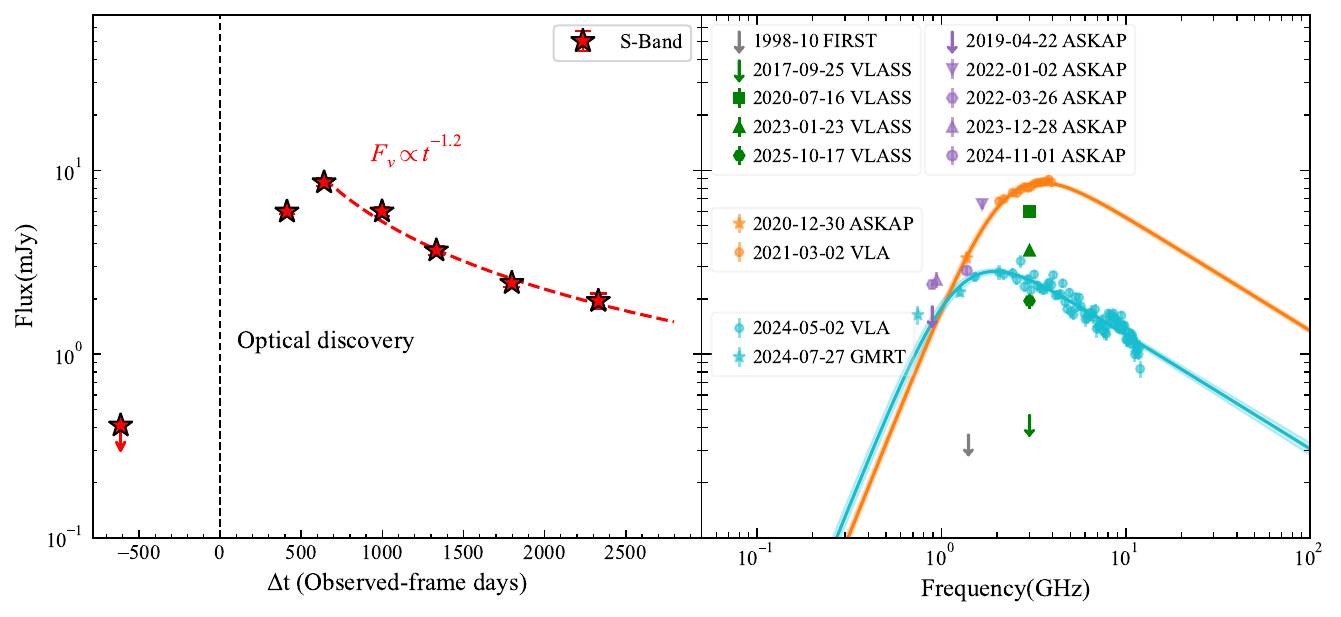}
\caption{
\label{fig:radio_sed}
Left: Radio light curve of AT2019ijn at S-Band (3 GHz), plotted relative to the the optical discovery date. The upper limit from VLASS epoch I is indicated by a red downward arrow, and the red dashed line represents a power law modeling of the data in the decay phase.
Right: Radio SEDs of AT2019ijn and the best-fitting models across two epochs. 
The solid lines show the best-fit models from our MCMC analysis, while the light shades regions represent the 1$\sigma$ confidence intervals derived from randomly selected posterior samples. Non-detections are marked with downward arrows, indicating 3$\sigma$ upper limits on the radio flux density.}
\end{figure*}

\subsection{Radio Flux and Luminosity Evolution}


The radio light curve of \src at 3 GHz (that is relatively well-sampled) is shown in Figure \ref{fig:radio_sed} (left). 
As we mentioned in Section 2.3.1, \src was undetected on 2017 Sep 25 during VLASS epoch I, about 613 days before the optical flare. It was first detected in the radio band during VLASS epoch II on 2020 Jul 16, with a 3 GHz flux brightening by a factor of $>$15. Such a transient brightening in the radio emission
is likely associated with the optical flare, as the latter has been found to occur on 2019 May 31, which is between the two VLASS observations. 
In comparison with the time of the optical discovery, 
we find a continuous flux rise from about 5.97 mJy to 8.61 mJy over a period of 229 days, 
followed by a slow decline lasting for at least 1690 days. 
Fitting the light curve with a power-law ($F_{\nu}\propto t^{-\alpha}$) in the decay phase, 
we find an index $\alpha=1.21\pm0.06$. 
Note that such a slow flux decay is consistent with the post-peak evolution of the radio emission from the jetted TDE Sw J1644+57 \citep{Cendes2021}. 

In Figure \ref{fig:radio_lc}, we show the radio luminosity evolution of \src at $\sim 1.3-3$ GHz, as well as a comparison to other 
radio-emitting transients at similar frequencies, including LFBOTs, jetted TDEs, GRBs and SNe.   
The 3 GHz radio luminosity of \src displays a slow increase to peak ($L_{\rm \nu, 3~GHz}= 2.0\times 10^{31}$ \ergshz) at $t_{\rm peak}=$ 641 days since the optical discovery, followed by a powerlaw decay (Figure \ref{fig:radio_sed}, left). 
At $\sim 1.3-1.7$ GHz, the radio luminosity evolution is similar, but peaks at a later time, with $t_{\rm peak}=$ 947 days. 
The high radio luminosity of \src is unprecedented in comparison with known SNe and LFBOTs. 
In addition to the luminosity, the radio evolution contrasts with the behavior of LFBOTs, which typically fade much more rapidly at rates of $\simgt t^{-3}$ \citep{Ho2023ApJ}. 
While the radio luminosity of \src is comparable to those of on-axis long GRBs from collapsars, the slow rise to peak at $t>600$ days post optical discovery is unusual and has never been observed before among GRBs with optical detections, including off-axis GRB candidates that typically peak within $t\simlt$100 days \citep{Li2025AT2023lcr,Perley2025, Srinivasaragavan2025}.
Its luminosity evolution appears to resemble the late-time emission from jetted TDEs, though peaks at a later time. 
The possible origins for the bright, slow-evolving radio emission will be discussed in Section 4.

\subsection{SED Evolution and Equipartition Analysis} 
\begin{deluxetable*}{ccccccccccc}
\centering
\tablewidth{0pt}
\setlength{\tabcolsep}{0.8mm}
\tablecaption{\textbf{Fitted Parameters of the Synchrotron Model for AT2019ijn} \label{tab:table2}}
\tablehead{
\colhead{} & \colhead{Epoch} & \colhead{$\delta t$}& \colhead{$\log R_{\rm eq}$} & \colhead{$\log E_{\rm eq}$} & \colhead{\bm{$\beta_{\rm eq} $}} & \colhead{\bm{$\log R$}}& \colhead{\bm{$\log E$}}& \colhead{\bm{$\beta$}} & \colhead{$\log \rm B$} & \colhead{$\log n_{\rm e}$}\\
\colhead{} & \colhead{} & \colhead{${\rm (days)}$} & \colhead{${\rm (cm)}$} & \colhead{${\rm (erg)}$} & \colhead{${(c)}$} & \colhead{${\rm (cm)}$} & \colhead{${\rm (erg)}$} & \colhead{${(c)}$} & \colhead{${\rm (G)}$} & \colhead{${\rm (cm^{-3})}$}
}
\setlength{\tabcolsep}{3mm}
\startdata
\multicolumn{11}{c}{$\epsilon_{\rm e} = 0.1, \epsilon_{\rm B} = 0.001$} \\
\hline
Spherical & I & 609.8 & $18.09_{-0.01}^{+0.01}$ & $51.95_{-0.01}^{+0.01}$ & $0.50$ & $17.99_{-0.01}^{+0.01}$ & $52.38_{-0.01}^{+0.01}$ & $0.44$ & $-1.41_{-0.01}^{+0.01}$ & $2.36_{-0.03}^{+0.02}$\\
& II & 1840.8 & $18.13_{-0.02}^{+0.02}$ & $51.64_{-0.05}^{+0.05}$ & $0.26$ & $18.03_{-0.02}^{+0.02}$ & $52.07_{-0.05}^{+0.05}$ & $0.22$ & $-1.63_{-0.03}^{+0.03}$ & $1.93_{-0.09}^{+0.09}$\\
Conical & I & 609.8 & $18.50_{-0.01}^{+0.01}$ & $52.12_{-0.01}^{+0.01}$ & $0.72$ & $18.40_{-0.01}^{+0.01}$ & $52.55_{-0.01}^{+0.01}$ & $0.67$ & $-1.51_{-0.01}^{+0.01}$ & $2.16_{-0.03}^{+0.02}$\\
& II & 1840.8 & $18.55_{-0.01}^{+0.02}$ & $51.81_{-0.05}^{+0.05}$ & $0.49$ & $18.45_{-0.02}^{+0.02}$ & $52.24_{-0.05}^{+0.05}$ & $0.43$ & $-1.73_{-0.03}^{+0.03}$ & $1.72_{-0.09}^{+0.09}$\\
\hline
\multicolumn{11}{c}{$\epsilon_{\rm e} = 0.1, \epsilon_{\rm B} = 0.01$} \\
\hline
Spherical & I & 609.8 & $18.09_{-0.01}^{+0.01}$ & $51.95_{-0.01}^{+0.01}$ & $0.50$ & $18.04_{-0.01}^{+0.01}$ & $52.06_{-0.01}^{+0.01}$ & $0.47$ & $-1.71_{-0.01}^{+0.01}$ & $1.80_{-0.03}^{+0.02}$\\
& II & 1840.8 & $18.13_{-0.02}^{+0.02}$ & $51.64_{-0.05}^{+0.05}$ & $0.26$ & $18.09_{-0.02}^{+0.02}$ & $51.76_{-0.05}^{+0.05}$ & $0.25$ & $-1.39_{-0.03}^{+0.03}$ & $1.37_{-0.09}^{+0.08}$\\
Conical & I & 609.8 & $18.50_{-0.01}^{+0.01}$ & $52.12_{-0.01}^{+0.01}$ & $0.72$ & $18.46_{-0.01}^{+0.01}$ & $52.23_{-0.01}^{+0.01}$ & $0.70$ & $-1.28_{-0.01}^{+0.01}$ & $1.59_{-0.03}^{+0.02}$\\
& II & 1840.8 & $18.55_{-0.02}^{+0.02}$ & $51.81_{-0.05}^{+0.05}$ & $0.49$ & $18.51_{-0.02}^{+0.02}$ & $51.93_{-0.05}^{+0.05}$ & $0.46$ & $-1.49_{-0.03}^{+0.03}$ & $1.16_{-0.09}^{+0.08}$\\
\hline
\multicolumn{11}{c}{$\epsilon_{\rm e} = 0.1, \epsilon_{\rm B} = 0.1$} \\
\hline
Spherical & I & 609.8 & $18.09_{-0.01}^{+0.01}$ & $51.95_{-0.01}^{+0.01}$ & $0.50$ & $18.10_{-0.01}^{+0.01}$ & $51.97_{-0.01}^{+0.01}$ & $0.50$ & $-0.94_{-0.01}^{+0.01}$ & $1.24_{-0.03}^{+0.02}$\\
& II & 1840.8 & $18.13_{-0.02}^{+0.02}$ & $51.64_{-0.05}^{+0.05}$ & $0.26$ & $18.15_{-0.02}^{+0.02}$ & $51.66_{-0.05}^{+0.05}$ & $0.27$ & $-1.16_{-0.03}^{+0.03}$ & $0.80_{-0.08}^{+0.08}$\\
Conical & I & 609.8 & $18.50_{-0.01}^{+0.01}$ & $52.12_{-0.01}^{+0.01}$ & $0.72$ & $18.52_{-0.01}^{+0.01}$ & $52.14_{-0.01}^{+0.01}$ & $0.73$ & $-1.04_{-0.01}^{+0.01}$ & $1.03_{-0.03}^{+0.02}$\\
& II & 1840.8 & $18.55_{-0.02}^{+0.02}$ & $51.81_{-0.05}^{+0.05}$ & $0.49$ & $18.56_{-0.02}^{+0.02}$ & $51.83_{-0.05}^{+0.05}$ & $0.49$ & $-1.26_{-0.03}^{+0.03}$ & $0.59_{-0.08}^{+0.08}$\\
\enddata
\end{deluxetable*}

Further insights into the radio properties can be obtained from the analysis of the radio spectral evolution as a function of time. 
Figure \ref{fig:radio_sed} (right) shows the radio SED in the 
$\sim0.6-10$ GHz and its time evolution over two epochs spanning by 1231 days.
For the first epoch, we used the data obtained from the ASKAP observations on 2020 Dec 30 and the VLA observations on 2021 March 2. 
For the second epoch, the SED was constructed using the quasi-simultaneous GMRT and VLA observations between 2024 April and July. 
It is clear that the SED exhibits a gradual shift to a lower peak flux density and frequency. 
To quantify the temporal evolution, we fit the SED with a synchrotron emission model in the context of an outflow 
expanding into the surrounding medium. 
The interaction of outflow with the surrounding medium leads to transient radio synchrotron emission owing to the acceleration of electrons and amplification of magnetic fields. 
This model, assuming the synchrotron self-absorption (SSA) to account for the low-frequency turnover, 
has been widely used to explain the late-time radio emission from various cosmic transients including 
SNe, GRBs, FBOTs and TDEs \citep[e.g.,][]{Chevalier1998, Granot2002, Berger2012, Margutti2019}.

In standard SSA model, the radio SED can be described by a smoothed broken power law (Equation (1) in \citet{Granot2002}):
\begin{equation}
F_\nu = F_{\nu_{\rm b, ext}} \left[ \left( \frac{\nu}{\nu_{\rm b}} \right)^{-s \beta_1} + \left( \frac{\nu}{\nu_{\rm b}} \right)^{-s \beta_2} \right]^{-1/s},
\label{equ1}
\end{equation}

where $F_{\nu_{\rm b, ext}}$ is the flux density at break frequency $\nu_{\rm b}$, $s$ is the sharpness parameter that smooths the transition between two asymptotic power-laws below and above $\nu_{\rm b}$.
We further adopt spectrum 2 from \citet{Granot2002}, assuming $\nu_{\rm m} \ll \nu_{\rm a} \ll \nu_{\rm c}$, where $\nu_{\rm m}$ is the characteristic synchrotron frequency of the emitting electrons with the least energy, $\nu_{\rm a}$ 
is the self-absorption frequency and $\nu_{\rm c}$ is the synchrotron cooling frequency.   
This is possible as the multi-frequency radio observations were performed at relatively late times ($\Delta t >600$ days), in which $\nu_m$ decreases more rapidly than $\nu_{\rm a}$ due to the adiabatic evolution of the shock. 
In this case, the spectral break occurs at
$\nu_{\rm a}$ and the spectral indices $\beta_1$ and $\beta_2$ correspond to the optically thick ($\nu<\nu_{\rm a}$) and thin regimes ($\nu>\nu_{\rm a}$), respectively, with $\beta_1= 5/2$ and $\beta_2 = -(p-1)/2$, where $p$ represents the power-law index of electron energy distribution ($dN(E)/dE = N_0E^{-p}$). 
The sharpness parameter is fixed at $s= 1.25 - 0.18p$. 
To fit the synchrotron SED using Equation \eqref{equ1}, we employed the MCMC fitting technique \citep[{\tt emcee},][]{Foreman-Mackey2013} with 128 walkers and 4000 steps ($\tau_{\rm max}= 41$).
We used top-hat priors and sufficient samples to get reliable posterior distributions. 
For the second epoch, the broadband SED allows us to infer the power-law index of the electrons, 
$p = 2.29\pm0.05$ and we applied this value to fit the SED obtained in the first epoch which has limited data points at higher frequencies. 
Figure \ref{fig:radio_sed} (right) shows the best-fitting SED models. 
From the SED fits we can determine the temporal evolution in the peak flux density and frequency, $F_{\rm p}$ and $\nu_{\rm p}$. 
We found that both $F_{\rm \nu_p}$ and $\rm \nu_p$ indeed decrease steadily with time, 
from $15.83\pm0.29$ mJy and $2.22\pm0.07$ GHz at $t=$ 610 days to $5.10\pm0.14$ mJy and $1.18\pm0.05$ GHz at $t=$ 1841 days. 

With the inferred values of $F_{\rm p}$ and $\rm \nu_{p}$, we adopted a standard equipartition analysis to derive the physical properties of the outflow and the radio-emitting region. 
Using the Equation(4, 5) of \citet{Goodwin2022} 
in the Newtonian regime as originally derived by \citet{Barniol2013}, we obtained the equipartition radius ($R_{\rm eq}$) and minimal total energy ($E_{\rm eq}=E_{\rm e}+E_{\rm B}$) in the conservative case of spherical geometry\footnote{$E_{\rm e}$ and $E_{\rm B}$ are the energy in relativistic electrons and magnetic field, and the total energy is minimized with respect to $R$ at $R_{\rm eq}$, with $E_{\rm B}=(6/11)E_{\rm e}$ \citep{Barniol2013}.}, with the area and volume filling factors of $f_{\rm A}=1$ and $f_{\rm V}=4/3$, respectively. 
We assumed the microphysical parameter of the energy fraction in electrons with a canonical value of $\epsilon_{\rm e}=0.1$ \citep[e.g.,][]{Barniol2013, Goodwin2022, Cendes2022}, considering that electrons are typically accelerated much less efficiently than protons \citep{Morlino2012}. 
We summarized the derived physical parameters with their associated uncertainties in Table \ref{tab:table2}.
The inferred $R_{\rm eq}$ shows a slight increase between epochs, from $1.23\pm0.03 \times 10^{18}$ to $1.35 \pm0.06\times 10^{18}$ cm. 
Such a slow expansion of the radio-emitting region suggests that the outflow might haven entered a significant deceleration phase.
Following \citet{Ho2019, Goodwin2022}, assuming the outflow was launched since the optical discovery, the average shock velocity $\beta_{\rm eq}$ can be estimated by $t \approx \frac{R_{\rm eq}(1 - \beta_{\rm eq})(1+z)}{\beta_{\rm eq} c}$. 
This yields a velocity of $\beta_{\rm eq}= 0.50$ in the first epoch, and $\beta_{\rm eq}= 0.26$ for the second, indicating that the outflow indeed has been decelerating.
The minimal shock energy 
that can explain the observed radio emission based on the equipartition analysis, decreases  
from $E_{\rm eq}=8.91\pm0.02\times10^{51}$ erg to $E_{\rm eq}=4.37\pm0.05 \times10^{51}$ erg over the same period.  
Note that although the multi-frequency radio observations for each epoch were not performed
quasi-simultaneously (separated by $\sim$2-3 months), the impact of potential intra-epoch
flux variability on the equipartition analysis is small. 
We tested that if interpolating flux at 1.4 GHz (L-band) to the epoch of 3 GHz (S-band) observation based on the L-band light curve, 
the changes on the derived SED parameters are at a level of 1.1\%-2.2\% for $F_{\rm p}$ and 4.1\%-8.5\% for $\nu_{\rm p}$, respectively. 
This corresponds to the changes in $R_{\rm eq}$ and $E_{\rm eq}$ by only a factor of 0.01-0.04 dex.

Since the total energy is minimized with $E_{\rm eq}$ at $R_{\rm eq}$, in which the electrons and the magnetic field are roughly at equipartition with $E_{\rm B} = \frac{6}{11} E_{\rm e}$, the ratio $\epsilon = \frac{\epsilon_{\rm B}}{\epsilon_{\rm e}}\frac{11}{6}$ parameterizes the deviation from equipartition.
While $E_{\rm eq}$ and $R_{\rm eq}$ remain consistent under a fixed $\epsilon_{\rm e}$, we corrected the radius and energy for the system by using the formulas $R = R_{\rm eq} \epsilon^{(1/17)}$ and $E = E_{\rm eq}((11/17)\epsilon^{(-6/17)} + (6/17)\epsilon^{(11/17)})$ \citep{Barniol2013} with varying fractions of the total energy in magnetic field ($\epsilon_{\rm B}$) from $10^{-3}$ to $10^{-1}$.
For $\epsilon_{\rm B}=10^{-3}$, the actual radius $R$ corresponding to the minimum energy deviates from $R_{\rm eq}$ by a multiplicative factor of 0.79 and the total nonthermal energy $E$ is greater than $E_{\rm eq}$ by a multiplicative factor of 2.68. 
For $\epsilon_{\rm B}=10^{-1}$, both radius and energy are slightly increased with $R=1.04 R_{\rm eq}$, $E=1.04 E_{\rm eq}$.
Based on the corrected $R$, we can also determine the magnetic field strength ($B$) and the ambient radiating electron density ($n_{\rm e}$) of the radio emitting region, using the Equation(8, 10, 12) of \citet{Goodwin2022} quoted from \citet{Barniol2013}. The results are summarized in Table \ref{tab:table2} for reference.  

Given the large average shock velocity and energy in the spherical case, we also consider the results for a collimated outflow, by adopting an half-opening angle of 30$^\circ$. 
This corresponds to the geometric factors of $f_{\rm A}=0.13$ and $f_{\rm V}=0.178$ \citep{Goodwin2022, Zhang2026,Zhou2026}. 
For the assumed collimated outflow, we found that equipartition radius and outflow energy are a factor of 2.57 and 1.48 times larger than in the spherical case, leading to a larger average velocity of $\beta_{\rm eq}=0.72$ and $\beta_{\rm eq}=0.49$ at $t=610$ and $t=1841$ days, respectively, under the same assumption that the outflow was launched at the time of optical discovery.
Note that the size of radio-emitting region depends on the opening angle of outflow. 
Smaller outflow opening angle is chosen, larger radius (and hence velocity) would be.

\section{Discussion} \label{sec:discussion}

The key observational properties of \src can be summarized as
follows: 

(1) \src was discovered as an optical transient on 2019 May 31, with 
a fast rise-time of a few days ($t_{\rm 1/2, rise}=3.7\pm0.2$ days). 
At $z=0.273$, \src has a high peak optical luminosity of $L_{\rm bol}=1.5\times10^{44}$ \ergs. 
Following the peak, the optical emission was evolving slowly, with a decay time-scale of $t_{\rm 1/2, decay}=46.3\pm1.5$ days. 
\src exhibits a persistent blue color ($g-r= -0.13$) over a period of at least 47 days since discovery. 
While the properties of fast rise-time, high optical peak luminosity, and blue color resemble those of LFBOTs \citep{Ho2023ApJ}, the slow decay time-scale is distinct. 

(2) The host galaxy is a dwarf galaxy with a stellar mass of $2\times 10^{8}$\msun~and sSFR of $7.9\times10^{-10}$ yr$^{-1}$, 
also comparable to those of LFBOTs. 
There is no evidence that the transient position deviates from the galaxy center, as we measured an offset of $\lesssim0.3^{\prime\prime}$ (3$\sigma$, $\lesssim1.3$ kpc in physical scale).

(3) A radio flare was clearly detected at the optical position of \srcs.
More interestingly, the peak time in the radio light curve at 3 GHz displays a delay of 641 days relative to the optical discovery. 
A similar delay in the radio light curve at $\sim 1.3-1.7$ GHz is also observed, which peaks even later. 
The radio luminosity is as high as $2 \times 10^{31}$ \ergshz, comparable to that of jetted TDEs at similar evolution epochs. 
An equipartition analysis of the radio spectral evolution suggests that a mildly relativistic outflow with velocity $\beta\simgt0.5$
and the minimum kinetic energy of $8.9\times10^{51}$ erg may have been launched.

\begin{figure}[t!]
\epsscale{1.2}
\plotone{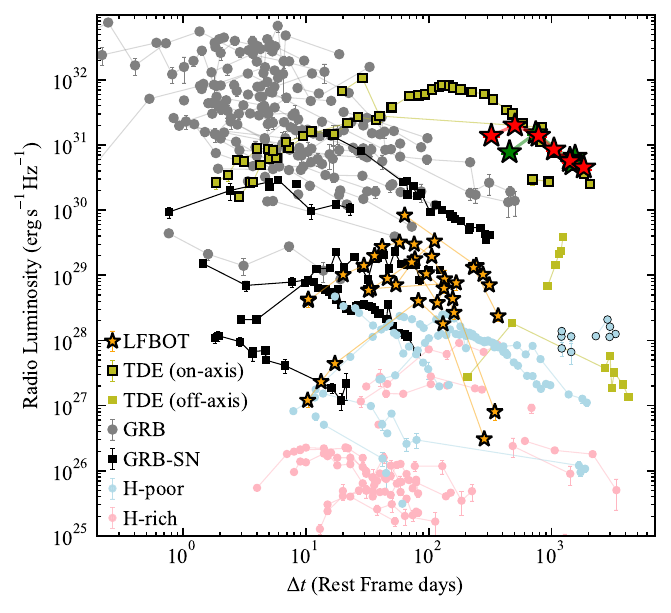}
\caption{
\label{fig:radio_lc}
The radio spectral luminosity evolution of AT2019ijn at S-band (3 GHz, red stars) and L-Band ($\sim$$1.3-1.7$ GHz, green stars). We also present light curves between 1-10 GHz of radio-luminous LFBOTs (orange stars), jetted-TDEs (olive-yellow squares), cosmological GRBs (gray dots), GRBs associated with SNe (black square), hydrogen-poor SNe (light blue dots, black edges denote represent radio-detected SLSNe), and hydrogen-rich SNe (light pink dots) for comparison.
While most of the archival SNe lie $<$ 50 Mpc, we do not correct the time to rest frame.
The archival GRBs are collected from \citet{Chandra2012, Levine2023} and references therein, include those associated with SNe, to which we add data from \citet{Finneran2024, Soderberg2010} and references therein.
SNe and SLSNe references include \citet{Bietenholz2021, Stroh2021, DeMarchi2022, Margutti2023, Eftekhari2021, Hatsukade2021APJL}, while LFBOTs are collected from \citet{Ho2019, Ho2020, Ho2022, Ho2023Nature, Margutti2019, Bietenholz2020, Coppejans2020, Bright2022, Yao2022, Chrimes2024AA, Nayana2025} and references therein.
We included Sw J1644+57, Sw J1112-82 Sw J2058+05, and AT2022cmc as on-axis jetted TDEs, while Arp 299-B AT1 and AT2018hyz as off-axis jetted TDEs \citep{Berger2012, Zauderer2013, Eftekhari2018, Cendes2021, Andreoni2022, Alexander2020, Cenko2012, Pasham2015, Brown2017,Mattila2018,Cendes2022}.}
\end{figure}

\begin{figure*}[t!]
\centering
\includegraphics[width=0.43\textwidth]{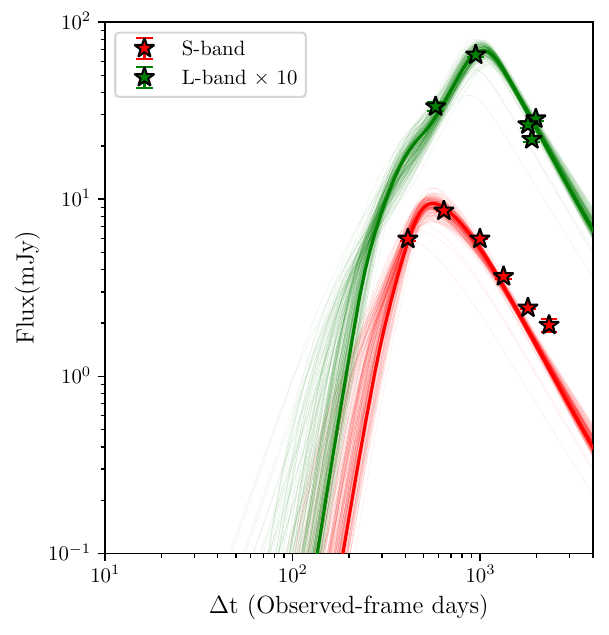}
\includegraphics[width=0.45\textwidth]{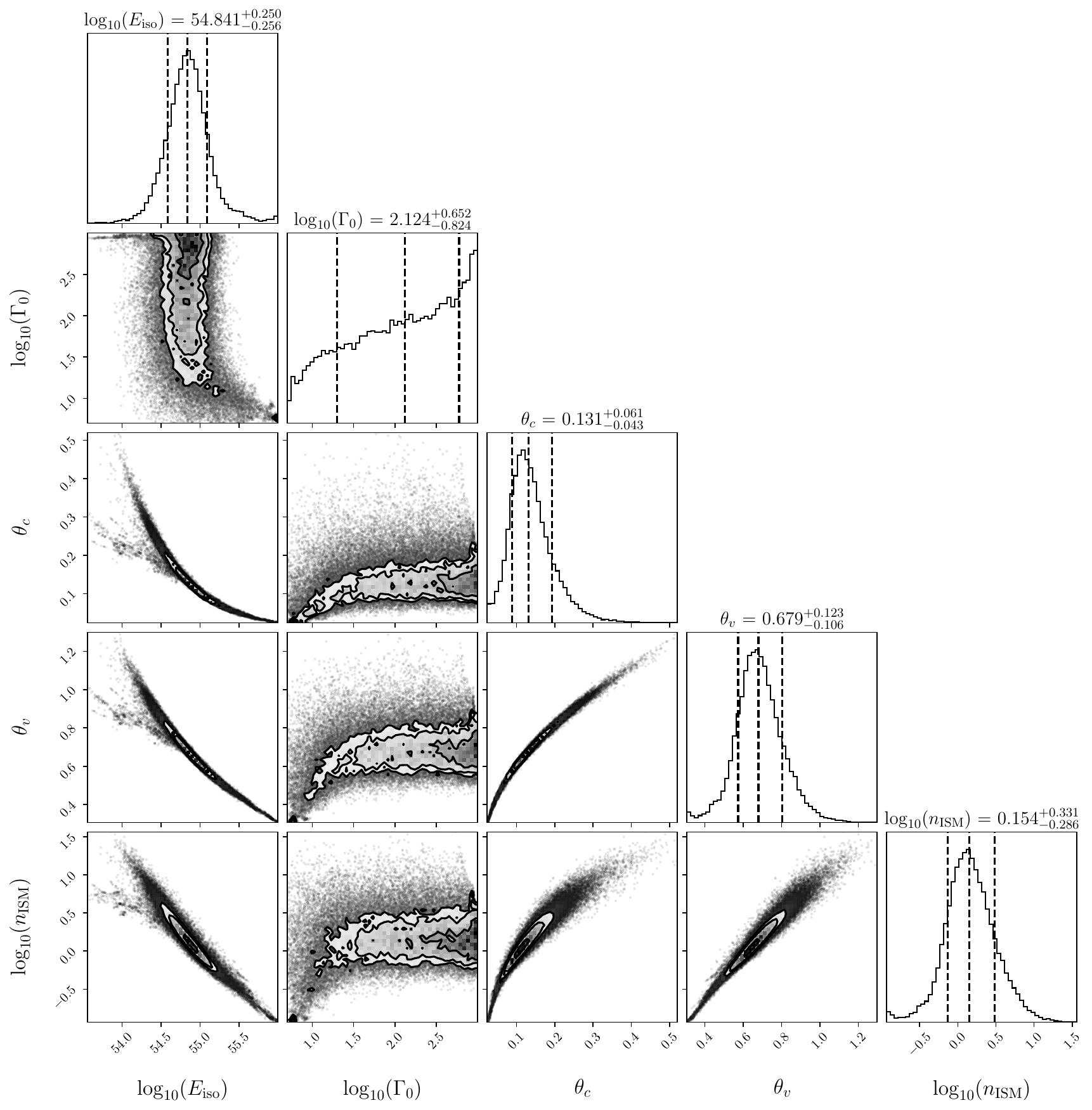}
\caption{
\label{fig:afterglow_fit}
Left: the results of modeling the S-band (3 GHz) and L-band ($\sim 1.3-1.7$ GHz in Table \ref{tab:table1}) data in the framework of radio afterglow emission from a jet interacting with surrounding medium. 
The colored lines represent model realizations: the light lines are based on 200 random samples from the MCMC chains, and the dark line represents the maximum likelihood.
Right: Corner plots showing the posterior distributions of the afterglow model parameters.
}
\end{figure*}

\subsection{On the Origin of the Exceptionally Bright Radio Emission} 
\label{subsection:off-axis jet}

As mentioned above, AT2019ijn displays luminous ($L_{\rm 3GHz}=2 \times 10^{31}$ \ergshz) and long-lasting ($t=412$ to $>2330$ days post discovery) radio emission (Figure \ref{fig:radio_lc}), standing out among known LFBOTs, which typically peak in $\sim$100 days with a luminosity $\lesssim10^{30}$ \ergshz, followed by a steep decay \citep{Ho2020, Coppejans2020, Chrimes2024AA, Nayana2025}. 
The exceptionally bright and long-lasting radio emission of \src cannot be explained by synchrotron radiation from the
supernova shock interaction with the CSM, which has a much lower peak luminosity 
of $\simlt10^{29}$ \ergshz at phases of years to a decade \citep{Nyland2020,Bietenholz2021,Stroh2021,Ibik2024}.
With reference to Figure \ref{fig:radio_lc}, the radio luminosity of AT2019ijn is comparable to those of cosmological long GRBs and jetted TDEs, suggesting that an energetic relativistic jet may have been launched.
However, \src differs from known on-axis GRBs in the observed timescales of radio evolution, as the luminosity for the latter typically decays by more than an order of magnitude in a few hundred days. 
In addition, no prompt radiation was observed in the $\gamma$-ray band at the time of optical discovery (Section \ref{subsection:High Energy}). 
\src could neither be associated with off-axis GRB candidates with optical detections, whose radio emission typically peaks within $t\simlt$100 days post optical discovery \citep{Li2025AT2023lcr,Perley2025, Srinivasaragavan2025}.
Conversely, its luminosity and timescale appear more similar to the jetted TDEs. 
The later peak time of the radio emission can be attributed to a relativistic jet viewed off-axis. 
In this case, the radio emission at early times will be strongly suppressed to the observer because of the relativistic beaming effect. 
As the jet decelerates, more radio emission spreads into the line of sight, leading to the light curve peaking at a later time \citep[e.g.,][]{Gupta2022,Li2025AT2023lcr,Perley2025}.

We explore the possibility of an off-axis jet model following the generalized equipartition analysis proposed in \citep{Matsumoto2023}. 
A critical parameter in their analysis is the apparent velocity ($\beta_{\rm eq, N}$) in Newtonian limits defined by $\beta_{\rm eq, N} = \frac{R_{\rm eq, N} (1+z)} {ct}$.
\citet{Beniamini2023} suggest that the critical value is $\beta_{\rm eq, N}=0.44$ for a more realistic maximum viewing angle of $\theta=\pi/2$, above which the relativistic off-axis solution can transition into
the Newtonian on-axis branch \citep{Matsumoto2023}. 
According to our equipartition analysis in the Newtonian regime (Section 3.3), the apparent velocity for \src approaches to 
$\beta_{\rm eq, N}=0.5$ by $\Delta t=$ 610 days, then drops to $\beta_{\rm eq, N}=0.26$ by $\Delta t=$ 1841 days.
Therefore, the radio emission of \srcs, which was observed to peak at $\Delta t=641$ days after the optical discovery, could 
have been produced by a relativistic jet viewed off-axis.

Assuming the radio emission originates from the relativistic jet interacting with the surrounding medium, we modeled the radio light curve at $\sim 1.3-1.7$ and 3 GHz using the open-source package {\tt VegasAfterglow} \citep{Wang2026} to derive physical constrains on an off-axis jet.
{\tt VegasAfterglow} is a high-performance framework designed for comprehensive, multi-wavelength afterglow modeling and robust MCMC parameter inference. 
We considered a top-hat jet with an initial Lorentz factor of $5<\Gamma_{0}<1000$, propagating through an ISM environment with a particle number density of $-2<log(n_{\rm ISM})<3$. The jet is characterized by an isotropic-equivalent kinetic energy of $50<log(E_{\rm iso})<56$ and half-opening angle of $0^\circ<\theta_{\rm c}<30^\circ$. The viewing angle spans in range $0^\circ<\theta_{\rm obs}<90^\circ$, covering both on- and off-axis orientations.
Due to the limited multi-frequency coverage, we fixed microphysical parameters at $p =2.5$, $\epsilon_{\rm e}=0.1$, while considering a range of $\epsilon_{\rm B}$ from $10^{-3}$ to $10^{-1}$. The MCMC fittings were performed using 16 walkers with $10^{6}$ steps ($\tau_{\rm max}= 2538$).
For the case of $\epsilon_{\rm B}=10^{-3}$, the best-fit model realizations  and corresponding posterior distributions of parameters are shown in Figure \ref{fig:afterglow_fit}.
While the afterglow model can reasonably describe the radio evolution in both L-band and S-band, there is a slight deviation for the latter at later times. 
Such a deviation could be due to the simplified model assumptions, such as a top-hat jet and/or a uniform ISM density. 
More detailed investigations on the effect of a structured jet or an inhomogeneous ISM are beyond the scope of current paper, and will be explored in future works if more late-time radio data are collected. 
The best-fit jet parameters from our afterglow modeling are $E_{\rm iso} = 6.9^{+5.4}_{-3.1} \times 10^{54}$ erg, $\theta_{\rm c} = 7.5^{+3.5}_{-2.5}\arcdeg$, $\theta_{\rm obs} = 38.9^{+7.0}_{-6.1}\arcdeg$, and $\Gamma_{0}\geq 19$, 
pointing to narrow, powerful jet with an off-axis viewing angle. 
Given the small jet opening angle, the beaming corrected energy can be calculated as $E_{\rm K} \sim 0.5 \theta_{0}^2 E_{\rm iso} = 5.9\times10^{52}$ erg \citep{Sfaradi2024}. 
Results for other values of $\epsilon_{\rm B}$ are summarized in Table \ref{tab:table3}. While assuming a higher magnetic field energy fraction of $\epsilon_{\rm B}=10^{-1}$ could lead to a decrease in the energy, the $E_{\rm iso}$ 
remains high at $6.7^{+1.7}_{-1.1}\times 10^{53}$ erg.
Such a high jet kinetic energy indicates that the embedded energy source is a compact object, such as a magnetar or an accreting black hole, 
both of which have been proposed to explain LFBOTs \citep[e.g.,][]{Margutti2019, Coppejans2020, Pasham2022, Zhang2022RAA, Ho2023Nature}. 
We will discuss the nature of the compact object powering \src in next Section. 

\subsection{Nature of the Powering Source} 

\subsubsection{Stellar-mass Compact Object?}
\label{subsection: Magnetar}

The first possibility we consider is that \src is powered by a stellar-mass compact object in the aftermath of a massive star collapse. 
In this scenario, there could be two possible energy sources: accretion for a black hole or rotational spindown for a neutron star. 
For a stellar-mass black hole ($M_{\rm BH}\sim10$\msun) where the optical emission is powered by accretion, the optical bolometric peak luminosity of $>10^{44}$ \ergs corresponds to an Eddington ratio of $L_{\rm engine}/L_{\rm Edd}>10^5$. 
This is too extreme for standard thin accretion disks, as it may exceed the Eddington luminosity by only a factor of $\sim$$10-100$ \citep{Begelman2002}. 
Alternatively, such an extreme super-Eddington accretion could be achieved if considering the stellar tidal disruption by a black hole. 
\citet{Metzger2022} proposed that a tidal disruption and hyper-accretion of a Wolf-Rayet star by a black hole binary companion 
(the accretion rate is $\sim$10 orders of magnitude larger than the Eddington rate) can account for the high optical luminosity of LFBOTs through outflow reprocessing. 
However, the model predicts a peak flattening of the light curve in the first a few days 
\citep[see Figures 2 and 3][]{Metzger2022}, which has not been observed in \srcs. 
More recently, \citet{Tsuna2025} proposed an alternative scenario for LFBOTs by invoking the tidal disruption of a main-sequence companion by a newborn black hole from a CCSNe explosion. 
The challenge for this model is the lack of evidence for the presence of a SNe in any LFBOTs reported thus far.  
Given its slow post-peak optical evolution that is distinct from known LFBOTs and 
exceptionally bright radio emission, explaining the powering source of \src in the context of stellar tidal disruption by a stellar-mass black hole binary companion remains further explored in theory, which is beyond the scope of this paper. 

On the other hand, the engine of \src could be a young magnetar (i.e., an extremely magnetized neutron star), 
where the optical luminosity is primarily powered by rotational energy loss due to spindown \citep{Margutti2019}. 
To explore 
the scenario of a millisecond magnetar as the central engine, 
we fitted the optical light curves with the magnetar model \citep{Nicholl2017} implemented in the Modular Open Source Fitter for Transients \citep[{\tt MOSFiT},][]{Guillochon2018}. 
To quantify the goodness of fit of the magnetar model, 
we adopted the Watanabe--Akaike information criteria \citep[WAIC;][]{Watanabe2010} \footnote{WAIC is a widely applicable Bayesian criterion for comparison between different models, and evaluated as $\rm WAIC=\langle logp_{n}\rangle-var(logp_n)$, where $\rm \langle logp_{n}\rangle$ is the mean of log likelihood score and $\rm var(logp_n)$ is its variance.}. The resulting WAIC is $90.1 \pm0.8$, 
indicating that the fittings are acceptable \citep{Nicholl2017}. 
Higher value of the WAIC score, better quality of the fit. 
The results are shown in Figure \ref{fig:mosfit} (upper panel).
The best-fit model suggests a magnetar spin period of $P_{\rm spin}=2.81_{-1.70}^{+1.23}$ ms, magnetic field strength of $B_{\perp}=0.32_{-0.17}^{+0.26} \times 10^{14}$ G, and an ejecta mass of $M_{\rm ej}=0.54_{-0.27}^{+0.49}$\msun. LFBOTs with multi-wavelength detections reach optical bolometric peak luminosities of 
$\simgt10^{44}$ \ergs, comparable to SLSNe (Figure \ref{fig:comparison}). 
It has been suggested that both SLSNe and FBOTs can be unified within the framework of magnetar engine, although their physical  parameters are different \citep{Liu2022}. For AT2019ijn, the ejecta mass $M_{\rm ej}$ is similar to that of FBOTs, but $P_{\rm spin}$ and $B_{\perp}$ are consistent with those of SLSNe.
This might explain why it has a fast-rising timescale similar to FBOTs, while it decays slowly, as observed in SLSNe. 
Therefore, while \src shares similarities with both FBOTs and SLSNe, the unique optical evolution properties clearly 
set it apart as a possibly new class of engine-driven transients.


 
\subsubsection{Jetted TDE Involving an IMBH?}
\label{subsection: TDE}
\begin{deluxetable}{cccc}
\centering
\tablewidth{0pt}
\setlength{\tabcolsep}{0.8mm}
\tablecaption{Fitted Parameters of the Afterglow Model for \src \label{tab:table3}}
\tablehead{
\colhead{Microphysical parameters} & \colhead{$\log E_{\rm iso}$} & \colhead{$\theta_{\rm c}$} & \colhead{$\theta_{\rm v}$}\\
\colhead{} & \colhead{${\rm (erg)}$} & \colhead{$\rm (rad)$} &\colhead{$\rm (rad)$}
}
\setlength{\tabcolsep}{3mm}
\startdata
$\epsilon_{\rm e} = 0.1, \epsilon_{\rm B} = 0.001$ & $54.84_{-0.26}^{+0.25}$ & $0.13_{-0.04}^{+0.06}$ & $0.68_{-0.11}^{+0.12}$\\
\hline
$\epsilon_{\rm e} = 0.1, \epsilon_{\rm B} = 0.01$ & $54.29_{-0.19}^{+0.18}$ & $0.16_{-0.04}^{+0.05}$ & $0.70_{-0.08}^{+0.10}$\\
\hline
$\epsilon_{\rm e} = 0.1, \epsilon_{\rm B} = 0.1$ & $53.82_{-0.08}^{+0.10}$ & $0.17_{-0.02}^{+0.02}$ & $0.68_{-0.04}^{+0.04}$\\
\enddata
\end{deluxetable}
So far, neither LFBOTs nor SLSNe have been observed with a clear evidence for relativistic jets. 
Although radio emission appears to be ubiquitous in LFBOTs, it can be best explained as originating from sub to mildly relativistic outflows. 
SN 2017ens and PTF10hgi are the only two SLSNe for which radio emission has been detected \citep{Eftekhari2019, Margutti2023}. 
However, due to the poor sampling in the radio light curve, the origin of the radio emission is not yet clear, 
possibly from a SNe shock interaction with the CSM, a magnetar wind nebula, or an off-axis jet. 
A notable example is SN 2011kl, a unique SNe with an intermediate luminosity between SLSNe and typical type Ic SNe. 
The association with the ultra-long GRB 111209A makes it a promising candidate SLSNe harboring a relativistic jet \citep{Greiner2015}. 
To reconcile with the optical properties of SN 2011kl in the framework of a magnetar model, an initial spin period of $\approx$$12$ ms for a magnetic field strength of (6-9)$\times10^{14}$ G is required.

If the central engine of AT2019ijn were indeed a magnetar, explaining its exceptionally bright radio emission would be challenging. The inferred isotropic-equivalent energy of the radio outflow reaches $\sim10^{54}$ erg, which is difficult to accommodate within a magnetar framework unless the energy release is strongly collimated into a very small solid angle. In addition, launching a relativistic magnetar jet requires several stringent conditions \citep{Yu2017,Desai2026}. These include millisecond initial rotation, ultra-strong magnetic fields, sufficiently low baryon loading along the polar regions enabled by multidimensional magnetohydrodynamic effects, effective jet collimation, and a favorable post-birth time window. The need to satisfy all of these requirements simultaneously suggests that a magnetar interpretation for AT2019ijn would be highly fine-tuned.



\begin{figure}[t!]
\epsscale{1.15}
\plotone{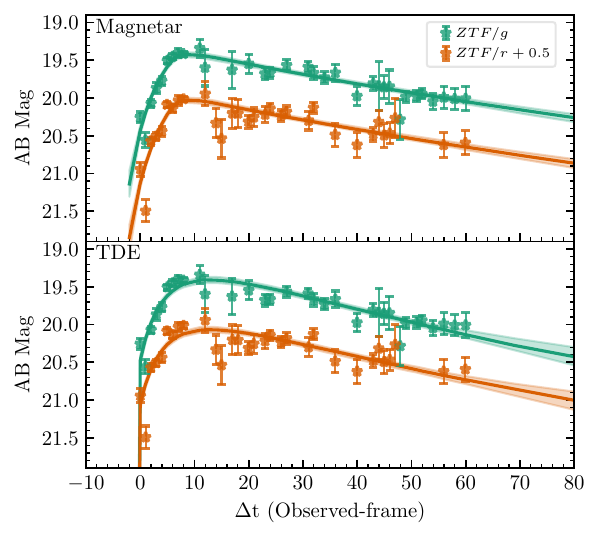}
\caption{
\label{fig:mosfit}
The ZTF light curves at $\textit{g}-$ and $\textit{r}-$bands, 
with the best-fit model realizations from {\tt MOSFiT}. 
The top panel shows the results from fittings with the magnetar model, while 
lower panel represents the TDE model. 
For clarity, the $\textit{r}-$band photometry has been arbitrarily shifted by adding $+0.5$ mag.
}
\end{figure}

As we mentioned in Section 3.2 (see also Figure \ref{fig:radio_lc}), 
the radio luminosity and evolution timescale of \src closely resemble jetted TDEs. 
In the optical bands, the rise timescale ($t_{\rm 1/2, rise}=3.7$ days) is shorter than most, if not all, optical TDEs discovered by ZTF (Figure \ref{fig:comparison}). 
The TDE AT2020neh has the shortest rise timescale in the ZTF sample \citep[$t_{\rm 1/2, rise}=6.4$ days,][]{Yao2023}, 
which has been interpreted as the tidal disruption of a main sequence star by an IMBH with mass $M_{\rm BH}\sim10^5$ $M_\odot$ \citep{Angus2022}. 
However, its peak absolute magnitude at $g$-band is $\approx-19.5$ mag, much less than that of \srcs. 
As shown in Figure \ref{fig:comparison}, the post-maximum evolution of \src is typical of other TDEs. 
Motivated by the similarity of \src to TDEs, 
we tested this possibility by fitting 
the observed multi-band optical light curves with the TDE model \citep{Mockler2019} using {\tt MOSFiT}.
Following \citet{Angus2022}, we did not impose the default cut of Eddington luminosity limit or the photosphere size limit during the fittings.
The results are displayed in Figure \ref{fig:mosfit} (lower panel), and the best-fit model invokes a black hole 
with mass log$M_{\rm BH} = 5.12_{-0.31}^{+0.28}$\msun~that is disrupting a relatively low-mass star of $m_\star = 0.22_{-0.14}^{+0.35}$\msun.
We found a WAIC value of $80.0 \pm0.6$, which falls into the WAIC range for a sample of optical TDEs \citep{Mockler2019}, 
indicating that the {\tt MOSFiT} results are reasonable.  
Such a relatively low black hole mass is consistent with that inferred for the TDE AT2020neh, though \src has a higher peak luminosity, indicating a higher accretion rate or radiation efficiency. 
In the context of TDEs, the fallback rate of the most bound stellar debris from the disrupted star 
depends on the black hole mass. 
For a main-sequence star disrupted, a lower mass for the disrupting black hole would result in a shorter time rising to peak luminosity, 
as $\Delta t \propto M_{\rm BH}^{1/2}$ \citep{Rees1988, Mockler2019, Angus2022}. 

AT2019ijn is hosted by a dwarf galaxy with stellar mass of $M_{\rm gal} = 2.04_{-1.23}^{+1.27}\times10^{8}$\msun.
By extrapolating the correlation between the stellar mass of a galaxy and the mass of central SMBH \citep{Kormendy2013, Reines2015, Greene2020} to the low-mass end, this $M_{\rm  gal}$ corresponds to a $M_{\rm BH}$ of $\sim10^5$\msun with a typical error of $\sim0.6-0.8$ dex, consistent with the mass estimated from the {\tt MOSFiT} modeling of the light curves.
This suggests that if \src is powered by TDE, an IMBH is required. 
Although the physical origin of LFBOTs remains unknown, one of the proposed models is a TDE involving an IMBH, 
such as AT2018cow \citep[e.g.,][]{Perley2019, Zhang2022RAA}, CSS161010 \citep{Coppejans2020, Gutierrez2024} and AT2024puz \citep{Somalwar2025}.    
This is supported by the detection of persistent, slowly fading UV and possibly X-ray emission from AT2018cow, 
which can be described by the late-time emission from an accretion disk formed after the TDE by a black hole with $M_{\rm BH}\sim10^{3}-10^4$\msun \citep[e.g.,][]{Migliori2024, Inkenhaag2025}. 
While most LFBOTs detected thus far have been found at a large offset from its host galaxy \citep[typically $\sim$1 kpc,][]{Chrimes2024MN}, 
the current optical imaging observations of \src have no sufficient positional accuracy 
to tell whether the transient is spatially consistent with the nucleus of its host or not.  



Only $\sim1\%$ of TDEs have been found to launch relativistic jets \citep{Andreoni2022}.
AT2022cmc, the first jetted TDE discovered in optical band, exhibits a fast rise and decay emission in the early phase and peaks brighter than $-$24 mag with a red color \citep{Andreoni2022}. The late-time evolution of thermal emission component appears to be consistent with the class of featureless and luminous TDEs \citep{Hammerstein2026}. The luminous, fast-fading red flare indicates the launching of an on-axis relativistic jet, following which the bright X-ray and radio emission are also detected within a few days \citep{Andreoni2022, Pasham2023}.
The precise mechanisms that regulate the production and evolution of jets in TDEs are poorly understood. 
By modeling the X-ray luminosity evolution, \citet{Eftekhari2024} suggested that jetted TDEs may preferentially form from lower mass black holes. 
More recently, the all-sky survey by Einstein Probe \citep{Yuan2025} has allowed to detect two fast and luminous X-ray transients, EP241021a and EP250702a, which could be associated with jetted TDEs involving IMBHs \citep[e.g.,][]{Shu2025, Li2025}. 
If this scenario is true, it would imply that the stellar tidal disruptions by IMBHs can 
launch relativistic jets.
\src may represent the first IMBH TDE coinciding with the center of a dwarf galaxy, for which a relativistic jet viewed off-axis is detected (Section 4.1). 
Future multiwavelength observations of LFBOTs similar to \src will help to uncover the nature of LFBOTs and 
the physical conditions for jet production, especially for IMBH accreting systems. 

\section{Conclusion} \label{sec:conclusion}


We report the discovery of a peculiar transient, \srcs, which displays unique emission properties in both optical and radio bands. 
The optical light curves show a fast rise time ($t_{\rm 1/2,rise} =3.7$ days), a persistently blue color (or $T_{\rm BB}\sim15000-16000$ K), and a high peak luminosity ($g-$band absolute magnitude of $-21.1$ mag), resembling those of LFBOTs.
However, it declines much slower than typical LFBOTs, with a decay time of $t_{\rm 1/2, decay}=46.3$ days, 
which is more consistent with those of SLSNe or TDEs.

The radio emission of \src is luminous and long-lasting, which peaks at $t=641$ days relative to the time of optical discovery, with a peak luminosity of $2\times10^{31}$ \ergshz. 
Such a radio luminosity is more than one order of magnitude higher than any other known LFBOTs and SLSNe, but comparable to events with relativistic jets like long GRBs and jetted TDEs. 
The long-sustained radio emission with a late-time peak disfavors a normal on-axis GRB powering the radio emission, unless it has an ultra-long duration. 
The radio emission from \src can be explained by an off-axis relativistic jet that spreads to our line of sight at late times, 
giving rise to the delayed peak in the light curve. 
In the context of this model, we find a powerful narrow jet with an opening angle of $\sim7-10\arcdeg$ and an isotropic equivalent kinetic energy of $\sim0.7-7\times10^{54}$ erg, which has a viewing angle of $\sim40\arcdeg$. 
Such a high jet kinetic energy, together with the luminous optical emission, suggests the presence of an active central engine. 
We have considered several scenarios as physical origins of \src 
and found a jetted TDE involving an IMBH is more favored. 

Given the faintness of the host galaxy, deep optical observations, such as the Hubble Space Telescope, are encouraged to better localize the galaxy's center. 
In addition, high-resolution VLBI observations are required to constrain the radio position at a milliarcsecond scale. 
This will enable to further explore the spatial coincidence between \src and the galaxy nucleus, which is crucial to confirm the nuclear origin.  
Our results suggest that 
AT2019ijn may represent a new class of relativistic optical transients, 
paving a new way for their statistical studies 
by combining optical and radio time-domain surveys. 

\acknowledgments{

{We thank the anonymous referee for the constructive suggestions and detailed comments, which helped to improve the clarity and quality of this paper.}
The data presented in this paper are based on observations
made with the Karl G. Jansky Very Large Array from the program VLA/20B-393, VLA/21B-322, and VLA/24A-078), 
and the Giant Metrewave Radio Telescope from the project ddtC366. 
We thank the staff of the VLA and GMRT that made these observations possible. 
The National Radio Astronomy Observatory is a facility of the
National Science Foundation operated under cooperative agreement
by Associated Universities, Inc.  
GMRT is run by the National Centre for Radio Astrophysics of the Tata Institute of Fundamental Research. 
The work is supported by 
the National Science Foundation of China (NSFC) through grant No. 12192220 and 12192221, 
{and National Key R\&D Program of China (No. 2025YFF0511101)}. 
X.S. acknowledges the science research grants from the China
Manned Space Project with CMS-CSST-2025-A07. 
This work has made use of data from the Asteroid Terrestrial-impact Last Alert System (ATLAS) project, 
and the ZTF forced-photometry service that was funded under the Heising-Simons Foundation grant \#12540303 (PI: Graham).
The Australian SKA Pathfinder is part of the Australia Telescope National Facility, which is managed by CSIRO. Operation of ASKAP is funded by the Australian Government with support from the National
Collaborative Research Infrastructure Strategy. This paper includes archived data obtained through the CSIRO ASKAP Science Data Archive, CASDA (http://data.csiro.au).  
}

\bibliography{ref}

@ARTICLE{Alexander2020,
       author = {{Alexander}, Kate D. and {van Velzen}, Sjoert and {Horesh}, Assaf and {Zauderer}, B. Ashley},
        title = "{Radio Properties of Tidal Disruption Events}",
      journal = {\ssr},
         year = 2020,
        month = jun,
       volume = {216},
       number = {5},
          eid = {81},
        pages = {81},
          doi = {10.1007/s11214-020-00702-w},
archivePrefix = {arXiv},
       eprint = {2006.01159},
 primaryClass = {astro-ph.HE},
       adsurl = {https://ui.adsabs.harvard.edu/abs/2020SSRv..216...81A}
}

@ARTICLE{Metzger2022,
       author = {{Metzger}, Brian D.},
        title = "{Luminous Fast Blue Optical Transients and Type Ibn/Icn SNe from Wolf-Rayet/Black Hole Mergers}",
      journal = {\apj},
         year = 2022,
        month = jun,
       volume = {932},
       number = {2},
          eid = {84},
        pages = {84},
          doi = {10.3847/1538-4357/ac6d59},
archivePrefix = {arXiv},
       eprint = {2203.04331},
 primaryClass = {astro-ph.HE},
       adsurl = {https://ui.adsabs.harvard.edu/abs/2022ApJ...932...84M}
}

@ARTICLE{Tsuna2025,
       author = {{Tsuna}, Daichi and {Lu}, Wenbin},
        title = "{Stellar Tidal Disruptions by Newborn Neutron Stars or Black Holes: A Mechanism for Hydrogen-poor (Super)luminous Supernovae and Fast Blue Optical Transients}",
      journal = {\apj},
         year = 2025,
        month = jun,
       volume = {986},
       number = {1},
          eid = {84},
        pages = {84},
          doi = {10.3847/1538-4357/add158},
archivePrefix = {arXiv},
       eprint = {2501.03316},
 primaryClass = {astro-ph.HE},
       adsurl = {https://ui.adsabs.harvard.edu/abs/2025ApJ...986...84T}
}

@ARTICLE{Pasham2023,
       author = {{Pasham}, Dheeraj R. and {Lucchini}, Matteo and {Laskar}, Tanmoy and {Gompertz}, Benjamin P. and {Srivastav}, Shubham and {Nicholl}, Matt and {Smartt}, Stephen J. and {Miller-Jones}, James C.~A. and {Alexander}, Kate D. and {Fender}, Rob and {Smith}, Graham P. and {Fulton}, M. and {Dewangan}, Gulab and {Gendreau}, Keith and {Coughlin}, Eric R. and {Rhodes}, Lauren and {Horesh}, Assaf and {van Velzen}, Sjoert and {Sfaradi}, Itai and {Guolo}, Muryel and {Castro Segura}, Noel and {Aamer}, Aysha and {Anderson}, Joseph P. and {Arcavi}, Iair and {Brennan}, Se{\'a}n J. and {Chambers}, Kenneth and {Charalampopoulos}, Panos and {Chen}, Ting-Wan and {Clocchiatti}, A. and {de Boer}, Thomas and {Dennefeld}, Michel and {Ferrara}, Elizabeth and {Galbany}, Llu{\'\i}s and {Gao}, Hua and {Gillanders}, James H. and {Goodwin}, Adelle and {Gromadzki}, Mariusz and {Huber}, M. and {Jonker}, Peter G. and {Joshi}, Manasvita and {Kara}, Erin and {Killestein}, Thomas L. and {Kosec}, Peter and {Kocevski}, Daniel and {Leloudas}, Giorgos and {Lin}, Chien-Cheng and {Margutti}, Raffaella and {Mattila}, Seppo and {Moore}, Thomas and {M{\"u}ller-Bravo}, Tom{\'a}s and {Ngeow}, Chow-Choong and {Oates}, Samantha and {Onori}, Francesca and {Pan}, Yen-Chen and {Perez-Torres}, Miguel and {Rani}, Priyanka and {Remillard}, Ronald and {Ridley}, Evan J. and {Schulze}, Steve and {Sheng}, Xinyue and {Shingles}, Luke and {Smith}, Ken W. and {Steiner}, James F. and {Wainscoat}, Richard and {Wevers}, Thomas and {Yang}, Sheng},
        title = "{The Birth of a Relativistic Jet Following the Disruption of a Star by a Cosmological Black Hole}",
      journal = {Nature Astronomy},
         year = 2023,
        month = jan,
       volume = {7},
        pages = {88-104},
          doi = {10.1038/s41550-022-01820-x},
archivePrefix = {arXiv},
       eprint = {2211.16537},
 primaryClass = {astro-ph.HE},
       adsurl = {https://ui.adsabs.harvard.edu/abs/2023NatAs...7...88P}
}

@ARTICLE{Begelman2002,
       author = {{Begelman}, Mitchell C.},
        title = "{Super-Eddington Fluxes from Thin Accretion Disks?}",
      journal = {\apjl},
         year = 2002,
        month = apr,
       volume = {568},
       number = {2},
        pages = {L97-L100},
          doi = {10.1086/340457},
archivePrefix = {arXiv},
       eprint = {astro-ph/0203030},
 primaryClass = {astro-ph},
       adsurl = {https://ui.adsabs.harvard.edu/abs/2002ApJ...568L..97B}
}

@ARTICLE{Eftekhari2024,
       author = {{Eftekhari}, T. and {Tchekhovskoy}, A. and {Alexander}, K.~D. and {Berger}, E. and {Chornock}, R. and {Laskar}, T. and {Margutti}, R. and {Yao}, Y. and {Cendes}, Y. and {Gomez}, S. and {Hajela}, A. and {Pasham}, D.~R.},
        title = "{Late-time X-Ray Observations of the Jetted Tidal Disruption Event AT2022cmc: The Relativistic Jet Shuts Off}",
      journal = {\apj},
         year = 2024,
        month = oct,
       volume = {974},
       number = {2},
          eid = {149},
        pages = {149},
          doi = {10.3847/1538-4357/ad72ea},
archivePrefix = {arXiv},
       eprint = {2404.10036},
 primaryClass = {astro-ph.HE},
       adsurl = {https://ui.adsabs.harvard.edu/abs/2024ApJ...974..149E}
}

@ARTICLE{Inkenhaag2025,
       author = {{Inkenhaag}, Anne and {Levan}, Andrew J. and {Mummery}, Andrew and {Jonker}, Peter G.},
        title = "{AT 2018cow at {\ensuremath{\sim}}5 years: additional evidence for a tidal disruption origin}",
      journal = {\mnras},
         year = 2025,
        month = nov,
       volume = {544},
       number = {1},
        pages = {L108-L112},
          doi = {10.1093/mnrasl/slaf107},
archivePrefix = {arXiv},
       eprint = {2510.08505},
 primaryClass = {astro-ph.HE},
       adsurl = {https://ui.adsabs.harvard.edu/abs/2025MNRAS.544L.108I}
}

@ARTICLE{Migliori2024,
       author = {{Migliori}, Giulia and {Margutti}, R. and {Metzger}, B.~D. and {Chornock}, R. and {Vignali}, C. and {Brethauer}, D. and {Coppejans}, D.~L. and {Maccarone}, T. and {Rivera Sandoval}, L. and {Bright}, J.~S. and {Laskar}, T. and {Milisavljevic}, D. and {Berger}, E. and {Nayana}, A.~J.},
        title = "{Roaring to Softly Whispering: X-Ray Emission after {\ensuremath{\sim}}3.7 yr at the Location of the Transient AT2018cow and Implications for Accretion-powered Scenarios}",
      journal = {\apjl},
         year = 2024,
        month = mar,
       volume = {963},
       number = {1},
          eid = {L24},
        pages = {L24},
          doi = {10.3847/2041-8213/ad2764},
       adsurl = {https://ui.adsabs.harvard.edu/abs/2024ApJ...963L..24M}
}

@ARTICLE{Pasham2022,
       author = {{Pasham}, Dheeraj R. and {Ho}, Wynn C.~G. and {Alston}, William and {Remillard}, Ronald and {Ng}, Mason and {Gendreau}, Keith and {Metzger}, Brian D. and {Altamirano}, Diego and {Chakrabarty}, Deepto and {Fabian}, Andrew and {Miller}, Jon and {Bult}, Peter and {Arzoumanian}, Zaven and {Steiner}, James F. and {Strohmayer}, Tod and {Tombesi}, Francesco and {Homan}, Jeroen and {Cackett}, Edward M. and {Harding}, Alice},
        title = "{Evidence for a compact object in the aftermath of the extragalactic transient AT2018cow}",
      journal = {Nature Astronomy},
         year = 2021,
        month = dec,
       volume = {6},
        pages = {249-258},
          doi = {10.1038/s41550-021-01524-8},
archivePrefix = {arXiv},
       eprint = {2112.04531},
 primaryClass = {astro-ph.HE},
       adsurl = {https://ui.adsabs.harvard.edu/abs/2022NatAs...6..249P}
}

@ARTICLE{Matsumoto2023,
       author = {{Matsumoto}, Tatsuya and {Piran}, Tsvi},
        title = "{Generalized equipartition method from an arbitrary viewing angle}",
      journal = {\mnras},
         year = 2023,
        month = jul,
       volume = {522},
       number = {3},
        pages = {4565-4576},
          doi = {10.1093/mnras/stad1269},
archivePrefix = {arXiv},
       eprint = {2211.10051},
 primaryClass = {astro-ph.HE},
       adsurl = {https://ui.adsabs.harvard.edu/abs/2023MNRAS.522.4565M}
}

@ARTICLE{Andreoni2022,
       author = {{Andreoni}, Igor and {Coughlin}, Michael W. and {Perley}, Daniel A. and {Yao}, Yuhan and {Lu}, Wenbin and {Cenko}, S. Bradley and {Kumar}, Harsh and {Anand}, Shreya and {Ho}, Anna Y.~Q. and {Kasliwal}, Mansi M. and {de Ugarte Postigo}, Antonio and {Sagu{\'e}s-Carracedo}, Ana and {Schulze}, Steve and {Kann}, D. Alexander and {Kulkarni}, S.~R. and {Sollerman}, Jesper and {Tanvir}, Nial and {Rest}, Armin and {Izzo}, Luca and {Somalwar}, Jean J. and {Kaplan}, David L. and {Ahumada}, Tom{\'a}s and {Anupama}, G.~C. and {Auchettl}, Katie and {Barway}, Sudhanshu and {Bellm}, Eric C. and {Bhalerao}, Varun and {Bloom}, Joshua S. and {Bremer}, Michael and {Bulla}, Mattia and {Burns}, Eric and {Campana}, Sergio and {Chandra}, Poonam and {Charalampopoulos}, Panos and {Cooke}, Jeff and {D'Elia}, Valerio and {Das}, Kaustav Kashyap and {Dobie}, Dougal and {Ag{\"u}{\'\i} Fern{\'a}ndez}, Jos{\'e} Feliciano and {Freeburn}, James and {Fremling}, Cristoffer and {Gezari}, Suvi and {Goode}, Simon and {Graham}, Matthew J. and {Hammerstein}, Erica and {Karambelkar}, Viraj R. and {Kilpatrick}, Charles D. and {Kool}, Erik C. and {Krips}, Melanie and {Laher}, Russ R. and {Leloudas}, Giorgos and {Levan}, Andrew and {Lundquist}, Michael J. and {Mahabal}, Ashish A. and {Medford}, Michael S. and {Miller}, M. Coleman and {M{\"o}ller}, Anais and {Mooley}, Kunal P. and {Nayana}, A.~J. and {Nir}, Guy and {Pang}, Peter T.~H. and {Paraskeva}, Emmy and {Perley}, Richard A. and {Petitpas}, Glen and {Pursiainen}, Miika and {Ravi}, Vikram and {Ridden-Harper}, Ryan and {Riddle}, Reed and {Rigault}, Mickael and {Rodriguez}, Antonio C. and {Rusholme}, Ben and {Sharma}, Yashvi and {Smith}, I.~A. and {Stein}, Robert D. and {Th{\"o}ne}, Christina and {Tohuvavohu}, Aaron and {Valdes}, Frank and {van Roestel}, Jan and {Vergani}, Susanna D. and {Wang}, Qinan and {Zhang}, Jielai},
        title = "{A very luminous jet from the disruption of a star by a massive black hole}",
      journal = {\nat},
         year = 2022,
        month = dec,
       volume = {612},
       number = {7940},
        pages = {430-434},
          doi = {10.1038/s41586-022-05465-8},
archivePrefix = {arXiv},
       eprint = {2211.16530},
 primaryClass = {astro-ph.HE},
       adsurl = {https://ui.adsabs.harvard.edu/abs/2022Natur.612..430A}
}

@ARTICLE{Angus2022,
       author = {{Angus}, C.~R. and {Baldassare}, V.~F. and {Mockler}, B. and {Foley}, R.~J. and {Ramirez-Ruiz}, E. and {Raimundo}, S.~I. and {French}, K.~D. and {Auchettl}, K. and {Pfister}, H. and {Gall}, C. and {Hjorth}, J. and {Drout}, M.~R. and {Alexander}, K.~D. and {Dimitriadis}, G. and {Hung}, T. and {Jones}, D.~O. and {Rest}, A. and {Siebert}, M.~R. and {Taggart}, K. and {Terreran}, G. and {Tinyanont}, S. and {Carroll}, C.~M. and {DeMarchi}, L. and {Earl}, N. and {Gagliano}, A. and {Izzo}, L. and {Villar}, V.~A. and {Zenati}, Y. and {Arendse}, N. and {Cold}, C. and {de Boer}, T.~J.~L. and {Chambers}, K.~C. and {Coulter}, D.~A. and {Khetan}, N. and {Lin}, C.~C. and {Magnier}, E.~A. and {Rojas-Bravo}, C. and {Wainscoat}, R.~J. and {Wojtak}, R.},
        title = "{A fast-rising tidal disruption event from a candidate intermediate-mass black hole}",
      journal = {Nature Astronomy},
         year = 2022,
        month = dec,
       volume = {6},
        pages = {1452-1463},
          doi = {10.1038/s41550-022-01811-y},
archivePrefix = {arXiv},
       eprint = {2209.00018},
 primaryClass = {astro-ph.HE},
       adsurl = {https://ui.adsabs.harvard.edu/abs/2022NatAs...6.1452A}
}

@ARTICLE{Barniol2013,
       author = {{Barniol Duran}, Rodolfo and {Nakar}, Ehud and {Piran}, Tsvi},
        title = "{Radius Constraints and Minimal Equipartition Energy of Relativistically Moving Synchrotron Sources}",
      journal = {\apj},
         year = 2013,
        month = jul,
       volume = {772},
       number = {1},
          eid = {78},
        pages = {78},
          doi = {10.1088/0004-637X/772/1/78},
archivePrefix = {arXiv},
       eprint = {1301.6759},
 primaryClass = {astro-ph.HE},
       adsurl = {https://ui.adsabs.harvard.edu/abs/2013ApJ...772...78B}
}

@ARTICLE{Bietenholz2021,
       author = {{Bietenholz}, M.~F. and {Bartel}, N. and {Argo}, M. and {Dua}, R. and {Ryder}, S. and {Soderberg}, A.},
        title = "{The Radio Luminosity-risetime Function of Core-collapse Supernovae}",
      journal = {\apj},
         year = 2021,
        month = feb,
       volume = {908},
       number = {1},
          eid = {75},
        pages = {75},
          doi = {10.3847/1538-4357/abccd9},
archivePrefix = {arXiv},
       eprint = {2011.11737},
 primaryClass = {astro-ph.HE},
       adsurl = {https://ui.adsabs.harvard.edu/abs/2021ApJ...908...75B}
}

@ARTICLE{Becker1995,
       author = {{Becker}, Robert H. and {White}, Richard L. and {Helfand}, David J.},
        title = "{The FIRST Survey: Faint Images of the Radio Sky at Twenty Centimeters}",
      journal = {\apj},
         year = 1995,
        month = sep,
       volume = {450},
        pages = {559},
          doi = {10.1086/176166},
       adsurl = {https://ui.adsabs.harvard.edu/abs/1995ApJ...450..559B}
}

@ARTICLE{Bellm2019,
       author = {{Bellm}, Eric C. and {Kulkarni}, Shrinivas R. and {Graham}, Matthew J. and {Dekany}, Richard and {Smith}, Roger M. and {Riddle}, Reed and {Masci}, Frank J. and {Helou}, George and {Prince}, Thomas A. and {Adams}, Scott M. and {Barbarino}, C. and {Barlow}, Tom and {Bauer}, James and {Beck}, Ron and {Belicki}, Justin and {Biswas}, Rahul and {Blagorodnova}, Nadejda and {Bodewits}, Dennis and {Bolin}, Bryce and {Brinnel}, Valery and {Brooke}, Tim and {Bue}, Brian and {Bulla}, Mattia and {Burruss}, Rick and {Cenko}, S. Bradley and {Chang}, Chan-Kao and {Connolly}, Andrew and {Coughlin}, Michael and {Cromer}, John and {Cunningham}, Virginia and {De}, Kishalay and {Delacroix}, Alex and {Desai}, Vandana and {Duev}, Dmitry A. and {Eadie}, Gwendolyn and {Farnham}, Tony L. and {Feeney}, Michael and {Feindt}, Ulrich and {Flynn}, David and {Franckowiak}, Anna and {Frederick}, S. and {Fremling}, C. and {Gal-Yam}, Avishay and {Gezari}, Suvi and {Giomi}, Matteo and {Goldstein}, Daniel A. and {Golkhou}, V. Zach and {Goobar}, Ariel and {Groom}, Steven and {Hacopians}, Eugean and {Hale}, David and {Henning}, John and {Ho}, Anna Y.~Q. and {Hover}, David and {Howell}, Justin and {Hung}, Tiara and {Huppenkothen}, Daniela and {Imel}, David and {Ip}, Wing-Huen and {Ivezi{\'c}}, {\v{Z}}eljko and {Jackson}, Edward and {Jones}, Lynne and {Juric}, Mario and {Kasliwal}, Mansi M. and {Kaspi}, S. and {Kaye}, Stephen and {Kelley}, Michael S.~P. and {Kowalski}, Marek and {Kramer}, Emily and {Kupfer}, Thomas and {Landry}, Walter and {Laher}, Russ R. and {Lee}, Chien-De and {Lin}, Hsing Wen and {Lin}, Zhong-Yi and {Lunnan}, Ragnhild and {Giomi}, Matteo and {Mahabal}, Ashish and {Mao}, Peter and {Miller}, Adam A. and {Monkewitz}, Serge and {Murphy}, Patrick and {Ngeow}, Chow-Choong and {Nordin}, Jakob and {Nugent}, Peter and {Ofek}, Eran and {Patterson}, Maria T. and {Penprase}, Bryan and {Porter}, Michael and {Rauch}, Ludwig and {Rebbapragada}, Umaa and {Reiley}, Dan and {Rigault}, Mickael and {Rodriguez}, Hector and {van Roestel}, Jan and {Rusholme}, Ben and {van Santen}, Jakob and {Schulze}, S. and {Shupe}, David L. and {Singer}, Leo P. and {Soumagnac}, Maayane T. and {Stein}, Robert and {Surace}, Jason and {Sollerman}, Jesper and {Szkody}, Paula and {Taddia}, F. and {Terek}, Scott and {Van Sistine}, Angela and {van Velzen}, Sjoert and {Vestrand}, W. Thomas and {Walters}, Richard and {Ward}, Charlotte and {Ye}, Quan-Zhi and {Yu}, Po-Chieh and {Yan}, Lin and {Zolkower}, Jeffry},
        title = "{The Zwicky Transient Facility: System Overview, Performance, and First Results}",
      journal = {\pasp},
         year = 2019,
        month = jan,
       volume = {131},
       number = {995},
        pages = {018002},
          doi = {10.1088/1538-3873/aaecbe},
archivePrefix = {arXiv},
       eprint = {1902.01932},
 primaryClass = {astro-ph.IM},
       adsurl = {https://ui.adsabs.harvard.edu/abs/2019PASP..131a8002B}
}

@ARTICLE{Bellm2019b,
       author = {{Bellm}, Eric C. and {Kulkarni}, Shrinivas R. and {Barlow}, Tom and {Feindt}, Ulrich and {Graham}, Matthew J. and {Goobar}, Ariel and {Kupfer}, Thomas and {Ngeow}, Chow-Choong and {Nugent}, Peter and {Ofek}, Eran and {Prince}, Thomas A. and {Riddle}, Reed and {Walters}, Richard and {Ye}, Quan-Zhi},
        title = "{The Zwicky Transient Facility: Surveys and Scheduler}",
      journal = {\pasp},
         year = 2019,
        month = jun,
       volume = {131},
       number = {1000},
        pages = {068003},
          doi = {10.1088/1538-3873/ab0c2a},
archivePrefix = {arXiv},
       eprint = {1905.02209},
 primaryClass = {astro-ph.IM},
       adsurl = {https://ui.adsabs.harvard.edu/abs/2019PASP..131f8003B}
}

@ARTICLE{Beniamini2023,
       author = {{Beniamini}, Paz and {Piran}, Tsvi and {Matsumoto}, Tatsuya},
        title = "{Swift J1644+57 as an off-axis Jet}",
      journal = {\mnras},
         year = 2023,
        month = sep,
       volume = {524},
       number = {1},
        pages = {1386-1395},
          doi = {10.1093/mnras/stad1950},
archivePrefix = {arXiv},
       eprint = {2305.06370},
 primaryClass = {astro-ph.HE},
       adsurl = {https://ui.adsabs.harvard.edu/abs/2023MNRAS.524.1386B}
}

@ARTICLE{Berger2012,
       author = {{Berger}, E. and {Zauderer}, A. and {Pooley}, G.~G. and {Soderberg}, A.~M. and {Sari}, R. and {Brunthaler}, A. and {Bietenholz}, M.~F.},
        title = "{Radio Monitoring of the Tidal Disruption Event Swift J164449.3+573451. I. Jet Energetics and the Pristine Parsec-scale Environment of a Supermassive Black Hole}",
      journal = {\apj},
         year = 2012,
        month = mar,
       volume = {748},
       number = {1},
          eid = {36},
        pages = {36},
          doi = {10.1088/0004-637X/748/1/36},
archivePrefix = {arXiv},
       eprint = {1112.1697},
 primaryClass = {astro-ph.HE},
       adsurl = {https://ui.adsabs.harvard.edu/abs/2012ApJ...748...36B}
}

@ARTICLE{Bietenholz2020,
       author = {{Bietenholz}, Michael F. and {Margutti}, Raffaella and {Coppejans}, Deanne and {Alexander}, Kate D. and {Argo}, Megan and {Bartel}, Norbert and {Eftekhari}, Tarraneh and {Milisavljevic}, Dan and {Terreran}, Giacomo and {Berger}, Edo},
        title = "{AT 2018cow VLBI: no long-lived relativistic outflow}",
      journal = {\mnras},
         year = 2020,
        month = feb,
       volume = {491},
       number = {4},
        pages = {4735-4741},
          doi = {10.1093/mnras/stz3249},
archivePrefix = {arXiv},
       eprint = {1911.08778},
 primaryClass = {astro-ph.HE},
       adsurl = {https://ui.adsabs.harvard.edu/abs/2020MNRAS.491.4735B}
}

@ARTICLE{Bright2022,
       author = {{Bright}, Joe S. and {Margutti}, Raffaella and {Matthews}, David and {Brethauer}, Daniel and {Coppejans}, Deanne and {Wieringa}, Mark H. and {Metzger}, Brian D. and {DeMarchi}, Lindsay and {Laskar}, Tanmoy and {Romero}, Charles and {Alexander}, Kate D. and {Horesh}, Assaf and {Migliori}, Giulia and {Chornock}, Ryan and {Berger}, E. and {Bietenholz}, Michael and {Devlin}, Mark J. and {Dicker}, Simon R. and {Jacobson-Gal{\'a}n}, W.~V. and {Mason}, Brian S. and {Milisavljevic}, Dan and {Motta}, Sara E. and {Mroczkowski}, Tony and {Ramirez-Ruiz}, Enrico and {Rhodes}, Lauren and {Sarazin}, Craig L. and {Sfaradi}, Itai and {Sievers}, Jonathan},
        title = "{Radio and X-Ray Observations of the Luminous Fast Blue Optical Transient AT 2020xnd}",
      journal = {\apj},
         year = 2022,
        month = feb,
       volume = {926},
       number = {2},
          eid = {112},
        pages = {112},
          doi = {10.3847/1538-4357/ac4506},
archivePrefix = {arXiv},
       eprint = {2110.05514},
 primaryClass = {astro-ph.HE},
       adsurl = {https://ui.adsabs.harvard.edu/abs/2022ApJ...926..112B}
}

@ARTICLE{Brown2017,
       author = {{Brown}, G.~C. and {Levan}, A.~J. and {Stanway}, E.~R. and {Kr{\"u}hler}, T. and {Tanvir}, N.~R. and {Davies}, L.~J.~M. and {Fruchter}, A. and {Cenko}, S.~B. and {Metzger}, B.~D.},
        title = "{Late-time observations of the relativistic tidal disruption flare candidate Swift J1112.2-8238}",
      journal = {\mnras},
         year = 2017,
        month = dec,
       volume = {472},
       number = {4},
        pages = {4469-4479},
          doi = {10.1093/mnras/stx2193},
archivePrefix = {arXiv},
       eprint = {1708.09668},
 primaryClass = {astro-ph.HE},
       adsurl = {https://ui.adsabs.harvard.edu/abs/2017MNRAS.472.4469B}
}

@ARTICLE{Bruzual2003,
       author = {{Bruzual}, G. and {Charlot}, S.},
        title = "{Stellar population synthesis at the resolution of 2003}",
      journal = {\mnras},
         year = 2003,
        month = oct,
       volume = {344},
       number = {4},
        pages = {1000-1028},
          doi = {10.1046/j.1365-8711.2003.06897.x},
archivePrefix = {arXiv},
       eprint = {astro-ph/0309134},
 primaryClass = {astro-ph},
       adsurl = {https://ui.adsabs.harvard.edu/abs/2003MNRAS.344.1000B}
}

@ARTICLE{CASATeam2022,
       author = {{CASA Team} and {Bean}, Ben and {Bhatnagar}, Sanjay and {Castro}, Sandra and {Donovan Meyer}, Jennifer and {Emonts}, Bjorn and {Garcia}, Enrique and {Garwood}, Robert and {Golap}, Kumar and {Gonzalez Villalba}, Justo and {Harris}, Pamela and {Hayashi}, Yohei and {Hoskins}, Josh and {Hsieh}, Mingyu and {Jagannathan}, Preshanth and {Kawasaki}, Wataru and {Keimpema}, Aard and {Kettenis}, Mark and {Lopez}, Jorge and {Marvil}, Joshua and {Masters}, Joseph and {McNichols}, Andrew and {Mehringer}, David and {Miel}, Renaud and {Moellenbrock}, George and {Montesino}, Federico and {Nakazato}, Takeshi and {Ott}, Juergen and {Petry}, Dirk and {Pokorny}, Martin and {Raba}, Ryan and {Rau}, Urvashi and {Schiebel}, Darrell and {Schweighart}, Neal and {Sekhar}, Srikrishna and {Shimada}, Kazuhiko and {Small}, Des and {Steeb}, Jan-Willem and {Sugimoto}, Kanako and {Suoranta}, Ville and {Tsutsumi}, Takahiro and {van Bemmel}, Ilse M. and {Verkouter}, Marjolein and {Wells}, Akeem and {Xiong}, Wei and {Szomoru}, Arpad and {Griffith}, Morgan and {Glendenning}, Brian and {Kern}, Jeff},
        title = "{CASA, the Common Astronomy Software Applications for Radio Astronomy}",
      journal = {\pasp},
         year = 2022,
        month = nov,
       volume = {134},
       number = {1041},
          eid = {114501},
        pages = {114501},
          doi = {10.1088/1538-3873/ac9642},
archivePrefix = {arXiv},
       eprint = {2210.02276},
 primaryClass = {astro-ph.IM},
       adsurl = {https://ui.adsabs.harvard.edu/abs/2022PASP..134k4501C}
}

@ARTICLE{Cendes2021,
       author = {{Cendes}, Y. and {Eftekhari}, T. and {Berger}, E. and {Polisensky}, E.},
        title = "{Radio Monitoring of the Tidal Disruption Event Swift J164449.3+573451. IV. Continued Fading and Non-relativistic Expansion}",
      journal = {\apj},
         year = 2021,
        month = feb,
       volume = {908},
       number = {2},
          eid = {125},
        pages = {125},
          doi = {10.3847/1538-4357/abd323},
archivePrefix = {arXiv},
       eprint = {2011.00074},
 primaryClass = {astro-ph.HE},
       adsurl = {https://ui.adsabs.harvard.edu/abs/2021ApJ...908..125C}
}

@ARTICLE{Cendes2022,
       author = {{Cendes}, Y. and {Berger}, E. and {Alexander}, K.~D. and {Gomez}, S. and {Hajela}, A. and {Chornock}, R. and {Laskar}, T. and {Margutti}, R. and {Metzger}, B. and {Bietenholz}, M.~F. and {Brethauer}, D. and {Wieringa}, M.~H.},
        title = "{A Mildly Relativistic Outflow Launched Two Years after Disruption in Tidal Disruption Event AT2018hyz}",
      journal = {\apj},
         year = 2022,
        month = oct,
       volume = {938},
       number = {1},
          eid = {28},
        pages = {28},
          doi = {10.3847/1538-4357/ac88d0},
archivePrefix = {arXiv},
       eprint = {2206.14297},
 primaryClass = {astro-ph.HE},
       adsurl = {https://ui.adsabs.harvard.edu/abs/2022ApJ...938...28C}
}

@ARTICLE{Cenko2012,
       author = {{Cenko}, S. Bradley and {Krimm}, Hans A. and {Horesh}, Assaf and {Rau}, Arne and {Frail}, Dale A. and {Kennea}, Jamie A. and {Levan}, Andrew J. and {Holland}, Stephen T. and {Butler}, Nathaniel R. and {Quimby}, Robert M. and {Bloom}, Joshua S. and {Filippenko}, Alexei V. and {Gal-Yam}, Avishay and {Greiner}, Jochen and {Kulkarni}, S.~R. and {Ofek}, Eran O. and {Olivares E.}, Felipe and {Schady}, Patricia and {Silverman}, Jeffrey M. and {Tanvir}, Nial R. and {Xu}, Dong},
        title = "{Swift J2058.4+0516: Discovery of a Possible Second Relativistic Tidal Disruption Flare?}",
      journal = {\apj},
         year = 2012,
        month = jul,
       volume = {753},
       number = {1},
          eid = {77},
        pages = {77},
          doi = {10.1088/0004-637X/753/1/77},
archivePrefix = {arXiv},
       eprint = {1107.5307},
 primaryClass = {astro-ph.HE},
       adsurl = {https://ui.adsabs.harvard.edu/abs/2012ApJ...753...77C}
}

@ARTICLE{Chabrier2003,
       author = {{Chabrier}, Gilles},
        title = "{Galactic Stellar and Substellar Initial Mass Function}",
      journal = {\pasp},
         year = 2003,
        month = jul,
       volume = {115},
       number = {809},
        pages = {763-795},
          doi = {10.1086/376392},
archivePrefix = {arXiv},
       eprint = {astro-ph/0304382},
 primaryClass = {astro-ph},
       adsurl = {https://ui.adsabs.harvard.edu/abs/2003PASP..115..763C}
}

@ARTICLE{Chambers2016,
       author = {{Chambers}, K.~C. and {Magnier}, E.~A. and {Metcalfe}, N. and {Flewelling}, H.~A. and {Huber}, M.~E. and {Waters}, C.~Z. and {Denneau}, L. and {Draper}, P.~W. and {Farrow}, D. and {Finkbeiner}, D.~P. and {Holmberg}, C. and {Koppenhoefer}, J. and {Price}, P.~A. and {Rest}, A. and {Saglia}, R.~P. and {Schlafly}, E.~F. and {Smartt}, S.~J. and {Sweeney}, W. and {Wainscoat}, R.~J. and {Burgett}, W.~S. and {Chastel}, S. and {Grav}, T. and {Heasley}, J.~N. and {Hodapp}, K.~W. and {Jedicke}, R. and {Kaiser}, N. and {Kudritzki}, R. -P. and {Luppino}, G.~A. and {Lupton}, R.~H. and {Monet}, D.~G. and {Morgan}, J.~S. and {Onaka}, P.~M. and {Shiao}, B. and {Stubbs}, C.~W. and {Tonry}, J.~L. and {White}, R. and {Ba{\~n}ados}, E. and {Bell}, E.~F. and {Bender}, R. and {Bernard}, E.~J. and {Boegner}, M. and {Boffi}, F. and {Botticella}, M.~T. and {Calamida}, A. and {Casertano}, S. and {Chen}, W. -P. and {Chen}, X. and {Cole}, S. and {Deacon}, N. and {Frenk}, C. and {Fitzsimmons}, A. and {Gezari}, S. and {Gibbs}, V. and {Goessl}, C. and {Goggia}, T. and {Gourgue}, R. and {Goldman}, B. and {Grant}, P. and {Grebel}, E.~K. and {Hambly}, N.~C. and {Hasinger}, G. and {Heavens}, A.~F. and {Heckman}, T.~M. and {Henderson}, R. and {Henning}, T. and {Holman}, M. and {Hopp}, U. and {Ip}, W. -H. and {Isani}, S. and {Jackson}, M. and {Keyes}, C.~D. and {Koekemoer}, A.~M. and {Kotak}, R. and {Le}, D. and {Liska}, D. and {Long}, K.~S. and {Lucey}, J.~R. and {Liu}, M. and {Martin}, N.~F. and {Masci}, G. and {McLean}, B. and {Mindel}, E. and {Misra}, P. and {Morganson}, E. and {Murphy}, D.~N.~A. and {Obaika}, A. and {Narayan}, G. and {Nieto-Santisteban}, M.~A. and {Norberg}, P. and {Peacock}, J.~A. and {Pier}, E.~A. and {Postman}, M. and {Primak}, N. and {Rae}, C. and {Rai}, A. and {Riess}, A. and {Riffeser}, A. and {Rix}, H.~W. and {R{\"o}ser}, S. and {Russel}, R. and {Rutz}, L. and {Schilbach}, E. and {Schultz}, A.~S.~B. and {Scolnic}, D. and {Strolger}, L. and {Szalay}, A. and {Seitz}, S. and {Small}, E. and {Smith}, K.~W. and {Soderblom}, D.~R. and {Taylor}, P. and {Thomson}, R. and {Taylor}, A.~N. and {Thakar}, A.~R. and {Thiel}, J. and {Thilker}, D. and {Unger}, D. and {Urata}, Y. and {Valenti}, J. and {Wagner}, J. and {Walder}, T. and {Walter}, F. and {Watters}, S.~P. and {Werner}, S. and {Wood-Vasey}, W.~M. and {Wyse}, R.},
        title = "{The Pan-STARRS1 Surveys}",
      journal = {arXiv e-prints},
         year = 2016,
        month = dec,
          eid = {arXiv:1612.05560},
        pages = {arXiv:1612.05560},
          doi = {10.48550/arXiv.1612.05560},
archivePrefix = {arXiv},
       eprint = {1612.05560},
 primaryClass = {astro-ph.IM},
       adsurl = {https://ui.adsabs.harvard.edu/abs/2016arXiv161205560C}
}

@ARTICLE{Chandra2012,
       author = {{Chandra}, Poonam and {Frail}, Dale A.},
        title = "{A Radio-selected Sample of Gamma-Ray Burst Afterglows}",
      journal = {\apj},
         year = 2012,
        month = feb,
       volume = {746},
       number = {2},
          eid = {156},
        pages = {156},
          doi = {10.1088/0004-637X/746/2/156},
archivePrefix = {arXiv},
       eprint = {1110.4124},
 primaryClass = {astro-ph.CO},
       adsurl = {https://ui.adsabs.harvard.edu/abs/2012ApJ...746..156C}
}

@ARTICLE{Chen2023,
       author = {{Chen}, Z.~H. and {Yan}, Lin and {Kangas}, T. and {Lunnan}, R. and {Schulze}, S. and {Sollerman}, J. and {Perley}, D.~A. and {Chen}, T. -W. and {Taggart}, K. and {Hinds}, K.~R. and {Gal-Yam}, A. and {Wang}, X.~F. and {Andreoni}, I. and {Bellm}, E. and {Bloom}, J.~S. and {Burdge}, K. and {Burgos}, A. and {Cook}, D. and {Dahiwale}, A. and {De}, K. and {Dekany}, R. and {Dugas}, A. and {Frederik}, S. and {Fremling}, C. and {Graham}, M. and {Hankins}, M. and {Ho}, A. and {Jencson}, J. and {Karambelkar}, V. and {Kasliwal}, M. and {Kulkarni}, S. and {Laher}, R. and {Rusholme}, B. and {Sharma}, Y. and {Taddia}, F. and {Tartaglia}, L. and {Thomas}, B.~P. and {Tzanidakis}, A. and {Van Roestel}, J. and {Walter}, R. and {Yang}, Y. and {Yao}, Y.~H. and {Yaron}, O.},
        title = "{The Hydrogen-poor Superluminous Supernovae from the Zwicky Transient Facility Phase I Survey. I. Light Curves and Measurements}",
      journal = {\apj},
         year = 2023,
        month = jan,
       volume = {943},
       number = {1},
          eid = {41},
        pages = {41},
          doi = {10.3847/1538-4357/aca161},
archivePrefix = {arXiv},
       eprint = {2202.02059},
 primaryClass = {astro-ph.HE},
       adsurl = {https://ui.adsabs.harvard.edu/abs/2023ApJ...943...41C}
}

@ARTICLE{Chevalier1998,
       author = {{Chevalier}, Roger A.},
        title = "{Synchrotron Self-Absorption in Radio Supernovae}",
      journal = {\apj},
         year = 1998,
        month = may,
       volume = {499},
       number = {2},
        pages = {810-819},
          doi = {10.1086/305676},
       adsurl = {https://ui.adsabs.harvard.edu/abs/1998ApJ...499..810C}
}

@ARTICLE{Chrimes2024MN,
       author = {{Chrimes}, A.~A. and {Jonker}, P.~G. and {Levan}, A.~J. and {Coppejans}, D.~L. and {Gaspari}, N. and {Gompertz}, B.~P. and {Groot}, P.~J. and {Malesani}, D.~B. and {Mummery}, A. and {Stanway}, E.~R. and {Wiersema}, K.},
        title = "{AT2023fhn (the Finch): a luminous fast blue optical transient at a large offset from its host galaxy}",
      journal = {\mnras},
         year = 2024,
        month = jan,
       volume = {527},
       number = {1},
        pages = {L47-L53},
          doi = {10.1093/mnrasl/slad145},
archivePrefix = {arXiv},
       eprint = {2307.01771},
 primaryClass = {astro-ph.HE},
       adsurl = {https://ui.adsabs.harvard.edu/abs/2024MNRAS.527L..47C}
}

@ARTICLE{Chrimes2024AA,
       author = {{Chrimes}, A.~A. and {Coppejans}, D.~L. and {Jonker}, P.~G. and {Levan}, A.~J. and {Groot}, P.~J. and {Mummery}, A. and {Stanway}, E.~R.},
        title = "{Multi-wavelength observations of the luminous fast blue optical transient AT 2023fhn: Up to {\ensuremath{\sim}}200 days post-explosion}",
      journal = {\aap},
         year = 2024,
        month = nov,
       volume = {691},
          eid = {A329},
        pages = {A329},
          doi = {10.1051/0004-6361/202451172},
archivePrefix = {arXiv},
       eprint = {2406.13821},
 primaryClass = {astro-ph.HE},
       adsurl = {https://ui.adsabs.harvard.edu/abs/2024A&A...691A.329C}
}

@ARTICLE{Cornwell2008,
       author = {{Cornwell}, T.~J. and {Golap}, K. and {Bhatnagar}, S.},
        title = "{The Noncoplanar Baselines Effect in Radio Interferometry: The W-Projection Algorithm}",
      journal = {IEEE Journal of Selected Topics in Signal Processing},
         year = 2008,
        month = nov,
       volume = {2},
       number = {5},
        pages = {647-657},
          doi = {10.1109/JSTSP.2008.2005290},
archivePrefix = {arXiv},
       eprint = {0807.4161},
 primaryClass = {astro-ph},
       adsurl = {https://ui.adsabs.harvard.edu/abs/2008ISTSP...2..647C}
}

@ARTICLE{Coppejans2020,
       author = {{Coppejans}, D.~L. and {Margutti}, R. and {Terreran}, G. and {Nayana}, A.~J. and {Coughlin}, E.~R. and {Laskar}, T. and {Alexander}, K.~D. and {Bietenholz}, M. and {Caprioli}, D. and {Chandra}, P. and {Drout}, M.~R. and {Frederiks}, D. and {Frohmaier}, C. and {Hurley}, K.~H. and {Kochanek}, C.~S. and {MacLeod}, M. and {Meisner}, A. and {Nugent}, P.~E. and {Ridnaia}, A. and {Sand}, D.~J. and {Svinkin}, D. and {Ward}, C. and {Yang}, S. and {Baldeschi}, A. and {Chilingarian}, I.~V. and {Dong}, Y. and {Esquivia}, C. and {Fong}, W. and {Guidorzi}, C. and {Lundqvist}, P. and {Milisavljevic}, D. and {Paterson}, K. and {Reichart}, D.~E. and {Shappee}, B. and {Stroh}, M.~C. and {Valenti}, S. and {Zauderer}, B.~A. and {Zhang}, B.},
        title = "{A Mildly Relativistic Outflow from the Energetic, Fast-rising Blue Optical Transient CSS161010 in a Dwarf Galaxy}",
      journal = {\apjl},
         year = 2020,
        month = may,
       volume = {895},
       number = {1},
          eid = {L23},
        pages = {L23},
          doi = {10.3847/2041-8213/ab8cc7},
archivePrefix = {arXiv},
       eprint = {2003.10503},
 primaryClass = {astro-ph.HE},
       adsurl = {https://ui.adsabs.harvard.edu/abs/2020ApJ...895L..23C}
}

@ARTICLE{DeMarchi2022,
       author = {{DeMarchi}, Lindsay and {Margutti}, R. and {Dittman}, J. and {Brunthaler}, A. and {Milisavljevic}, D. and {Bietenholz}, Michael F. and {Stauffer}, C. and {Brethauer}, D. and {Coppejans}, D. and {Auchettl}, K. and {Alexander}, K.~D. and {Kilpatrick}, C.~D. and {Bright}, Joe S. and {Kelley}, L.~Z. and {Stroh}, Michael C. and {Jacobson-Gal{\'a}n}, W.~V.},
        title = "{Radio Analysis of SN2004C Reveals an Unusual CSM Density Profile as a Harbinger of Core Collapse}",
      journal = {\apj},
         year = 2022,
        month = oct,
       volume = {938},
       number = {1},
          eid = {84},
        pages = {84},
          doi = {10.3847/1538-4357/ac8c26},
archivePrefix = {arXiv},
       eprint = {2203.07388},
 primaryClass = {astro-ph.HE},
       adsurl = {https://ui.adsabs.harvard.edu/abs/2022ApJ...938...84D}
}

@ARTICLE{Dey2019,
       author = {{Dey}, Arjun and {Schlegel}, David J. and {Lang}, Dustin and {Blum}, Robert and {Burleigh}, Kaylan and {Fan}, Xiaohui and {Findlay}, Joseph R. and {Finkbeiner}, Doug and {Herrera}, David and {Juneau}, St{\'e}phanie and {Landriau}, Martin and {Levi}, Michael and {McGreer}, Ian and {Meisner}, Aaron and {Myers}, Adam D. and {Moustakas}, John and {Nugent}, Peter and {Patej}, Anna and {Schlafly}, Edward F. and {Walker}, Alistair R. and {Valdes}, Francisco and {Weaver}, Benjamin A. and {Y{\`e}che}, Christophe and {Zou}, Hu and {Zhou}, Xu and {Abareshi}, Behzad and {Abbott}, T.~M.~C. and {Abolfathi}, Bela and {Aguilera}, C. and {Alam}, Shadab and {Allen}, Lori and {Alvarez}, A. and {Annis}, James and {Ansarinejad}, Behzad and {Aubert}, Marie and {Beechert}, Jacqueline and {Bell}, Eric F. and {BenZvi}, Segev Y. and {Beutler}, Florian and {Bielby}, Richard M. and {Bolton}, Adam S. and {Brice{\~n}o}, C{\'e}sar and {Buckley-Geer}, Elizabeth J. and {Butler}, Karen and {Calamida}, Annalisa and {Carlberg}, Raymond G. and {Carter}, Paul and {Casas}, Ricard and {Castander}, Francisco J. and {Choi}, Yumi and {Comparat}, Johan and {Cukanovaite}, Elena and {Delubac}, Timoth{\'e}e and {DeVries}, Kaitlin and {Dey}, Sharmila and {Dhungana}, Govinda and {Dickinson}, Mark and {Ding}, Zhejie and {Donaldson}, John B. and {Duan}, Yutong and {Duckworth}, Christopher J. and {Eftekharzadeh}, Sarah and {Eisenstein}, Daniel J. and {Etourneau}, Thomas and {Fagrelius}, Parker A. and {Farihi}, Jay and {Fitzpatrick}, Mike and {Font-Ribera}, Andreu and {Fulmer}, Leah and {G{\"a}nsicke}, Boris T. and {Gaztanaga}, Enrique and {George}, Koshy and {Gerdes}, David W. and {Gontcho}, Satya Gontcho A. and {Gorgoni}, Claudio and {Green}, Gregory and {Guy}, Julien and {Harmer}, Diane and {Hernandez}, M. and {Honscheid}, Klaus and {Huang}, Lijuan Wendy and {James}, David J. and {Jannuzi}, Buell T. and {Jiang}, Linhua and {Joyce}, Richard and {Karcher}, Armin and {Karkar}, Sonia and {Kehoe}, Robert and {Kneib}, Jean-Paul and {Kueter-Young}, Andrea and {Lan}, Ting-Wen and {Lauer}, Tod R. and {Le Guillou}, Laurent and {Le Van Suu}, Auguste and {Lee}, Jae Hyeon and {Lesser}, Michael and {Perreault Levasseur}, Laurence and {Li}, Ting S. and {Mann}, Justin L. and {Marshall}, Robert and {Mart{\'\i}nez-V{\'a}zquez}, C.~E. and {Martini}, Paul and {du Mas des Bourboux}, H{\'e}lion and {McManus}, Sean and {Meier}, Tobias Gabriel and {M{\'e}nard}, Brice and {Metcalfe}, Nigel and {Mu{\~n}oz-Guti{\'e}rrez}, Andrea and {Najita}, Joan and {Napier}, Kevin and {Narayan}, Gautham and {Newman}, Jeffrey A. and {Nie}, Jundan and {Nord}, Brian and {Norman}, Dara J. and {Olsen}, Knut A.~G. and {Paat}, Anthony and {Palanque-Delabrouille}, Nathalie and {Peng}, Xiyan and {Poppett}, Claire L. and {Poremba}, Megan R. and {Prakash}, Abhishek and {Rabinowitz}, David and {Raichoor}, Anand and {Rezaie}, Mehdi and {Robertson}, A.~N. and {Roe}, Natalie A. and {Ross}, Ashley J. and {Ross}, Nicholas P. and {Rudnick}, Gregory and {Safonova}, Sasha and {Saha}, Abhijit and {S{\'a}nchez}, F. Javier and {Savary}, Elodie and {Schweiker}, Heidi and {Scott}, Adam and {Seo}, Hee-Jong and {Shan}, Huanyuan and {Silva}, David R. and {Slepian}, Zachary and {Soto}, Christian and {Sprayberry}, David and {Staten}, Ryan and {Stillman}, Coley M. and {Stupak}, Robert J. and {Summers}, David L. and {Sien Tie}, Suk and {Tirado}, H. and {Vargas-Maga{\~n}a}, Mariana and {Vivas}, A. Katherina and {Wechsler}, Risa H. and {Williams}, Doug and {Yang}, Jinyi and {Yang}, Qian and {Yapici}, Tolga and {Zaritsky}, Dennis and {Zenteno}, A. and {Zhang}, Kai and {Zhang}, Tianmeng and {Zhou}, Rongpu and {Zhou}, Zhimin},
        title = "{Overview of the DESI Legacy Imaging Surveys}",
      journal = {\aj},
         year = 2019,
        month = may,
       volume = {157},
       number = {5},
          eid = {168},
        pages = {168},
          doi = {10.3847/1538-3881/ab089d},
archivePrefix = {arXiv},
       eprint = {1804.08657},
 primaryClass = {astro-ph.IM},
       adsurl = {https://ui.adsabs.harvard.edu/abs/2019AJ....157..168D}
}

@ARTICLE{Dong2021,
       author = {{Dong}, D.~Z. and {Hallinan}, G. and {Nakar}, E. and {Ho}, A.~Y.~Q. and {Hughes}, A.~K. and {Hotokezaka}, K. and {Myers}, S.~T. and {De}, K. and {Mooley}, K.~P. and {Ravi}, V. and {Horesh}, A. and {Kasliwal}, M.~M. and {Kulkarni}, S.~R.},
        title = "{A transient radio source consistent with a merger-triggered core collapse supernova}",
      journal = {Science},
         year = 2021,
        month = sep,
       volume = {373},
       number = {6559},
        pages = {1125-1129},
          doi = {10.1126/science.abg6037},
archivePrefix = {arXiv},
       eprint = {2109.01752},
 primaryClass = {astro-ph.HE},
       adsurl = {https://ui.adsabs.harvard.edu/abs/2021Sci...373.1125D}
}

@ARTICLE{Drout2014,
       author = {{Drout}, M.~R. and {Chornock}, R. and {Soderberg}, A.~M. and {Sanders}, N.~E. and {McKinnon}, R. and {Rest}, A. and {Foley}, R.~J. and {Milisavljevic}, D. and {Margutti}, R. and {Berger}, E. and {Calkins}, M. and {Fong}, W. and {Gezari}, S. and {Huber}, M.~E. and {Kankare}, E. and {Kirshner}, R.~P. and {Leibler}, C. and {Lunnan}, R. and {Mattila}, S. and {Marion}, G.~H. and {Narayan}, G. and {Riess}, A.~G. and {Roth}, K.~C. and {Scolnic}, D. and {Smartt}, S.~J. and {Tonry}, J.~L. and {Burgett}, W.~S. and {Chambers}, K.~C. and {Hodapp}, K.~W. and {Jedicke}, R. and {Kaiser}, N. and {Magnier}, E.~A. and {Metcalfe}, N. and {Morgan}, J.~S. and {Price}, P.~A. and {Waters}, C.},
        title = "{Rapidly Evolving and Luminous Transients from Pan-STARRS1}",
      journal = {\apj},
         year = 2014,
        month = oct,
       volume = {794},
       number = {1},
          eid = {23},
        pages = {23},
          doi = {10.1088/0004-637X/794/1/23},
archivePrefix = {arXiv},
       eprint = {1405.3668},
 primaryClass = {astro-ph.HE},
       adsurl = {https://ui.adsabs.harvard.edu/abs/2014ApJ...794...23D}
}

@ARTICLE{Eftekhari2018,
       author = {{Eftekhari}, T. and {Berger}, E. and {Zauderer}, B.~A. and {Margutti}, R. and {Alexander}, K.~D.},
        title = "{Radio Monitoring of the Tidal Disruption Event Swift J164449.3+573451. III. Late-time Jet Energetics and a Deviation from Equipartition}",
      journal = {\apj},
         year = 2018,
        month = feb,
       volume = {854},
       number = {2},
          eid = {86},
        pages = {86},
          doi = {10.3847/1538-4357/aaa8e0},
archivePrefix = {arXiv},
       eprint = {1710.07289},
 primaryClass = {astro-ph.HE},
       adsurl = {https://ui.adsabs.harvard.edu/abs/2018ApJ...854...86E}
}

@ARTICLE{Eftekhari2019,
       author = {{Eftekhari}, T. and {Berger}, E. and {Margalit}, B. and {Blanchard}, P.~K. and {Patton}, L. and {Demorest}, P. and {Williams}, P.~K.~G. and {Chatterjee}, S. and {Cordes}, J.~M. and {Lunnan}, R. and {Metzger}, B.~D. and {Nicholl}, M.},
        title = "{A Radio Source Coincident with the Superluminous Supernova PTF10hgi: Evidence for a Central Engine and an Analog of the Repeating FRB 121102?}",
      journal = {\apjl},
         year = 2019,
        month = may,
       volume = {876},
       number = {1},
          eid = {L10},
        pages = {L10},
          doi = {10.3847/2041-8213/ab18a5},
archivePrefix = {arXiv},
       eprint = {1901.10479},
 primaryClass = {astro-ph.HE},
       adsurl = {https://ui.adsabs.harvard.edu/abs/2019ApJ...876L..10E}
}

@ARTICLE{Eftekhari2021,
       author = {{Eftekhari}, T. and {Margalit}, B. and {Omand}, C.~M.~B. and {Berger}, E. and {Blanchard}, P.~K. and {Demorest}, P. and {Metzger}, B.~D. and {Murase}, K. and {Nicholl}, M. and {Villar}, V.~A. and {Williams}, P.~K.~G. and {Alexander}, K.~D. and {Chatterjee}, S. and {Coppejans}, D.~L. and {Cordes}, J.~M. and {Gomez}, S. and {Hosseinzadeh}, G. and {Hsu}, B. and {Kashiyama}, K. and {Margutti}, R. and {Yin}, Y.},
        title = "{Late-time Radio and Millimeter Observations of Superluminous Supernovae and Long Gamma-Ray Bursts: Implications for Central Engines, Fast Radio Bursts, and Obscured Star Formation}",
      journal = {\apj},
         year = 2021,
        month = may,
       volume = {912},
       number = {1},
          eid = {21},
        pages = {21},
          doi = {10.3847/1538-4357/abe9b8},
archivePrefix = {arXiv},
       eprint = {2010.06612},
 primaryClass = {astro-ph.HE},
       adsurl = {https://ui.adsabs.harvard.edu/abs/2021ApJ...912...21E}
}

@ARTICLE{Finneran2024,
       author = {{Finneran}, Gabriel and {Cotter}, Laura and {Martin-Carrillo}, Antonio},
        title = "{The GRBSN webtool: An open-source repository for gamma-ray burst-supernova associations}",
      journal = {arXiv e-prints},
         year = 2024,
        month = nov,
          eid = {arXiv:2411.08866},
        pages = {arXiv:2411.08866},
          doi = {10.48550/arXiv.2411.08866},
archivePrefix = {arXiv},
       eprint = {2411.08866},
 primaryClass = {astro-ph.HE},
       adsurl = {https://ui.adsabs.harvard.edu/abs/2024arXiv241108866F}
}

@ARTICLE{Fitzpatrick1999,
       author = {{Fitzpatrick}, Edward L.},
        title = "{Correcting for the Effects of Interstellar Extinction}",
      journal = {\pasp},
         year = 1999,
        month = jan,
       volume = {111},
       number = {755},
        pages = {63-75},
          doi = {10.1086/316293},
archivePrefix = {arXiv},
       eprint = {astro-ph/9809387},
 primaryClass = {astro-ph},
       adsurl = {https://ui.adsabs.harvard.edu/abs/1999PASP..111...63F}
}

@ARTICLE{Foreman-Mackey2013,
       author = {{Foreman-Mackey}, Daniel and {Hogg}, David W. and {Lang}, Dustin and {Goodman}, Jonathan},
        title = "{emcee: The MCMC Hammer}",
      journal = {\pasp},
         year = 2013,
        month = mar,
       volume = {125},
       number = {925},
        pages = {306},
          doi = {10.1086/670067},
archivePrefix = {arXiv},
       eprint = {1202.3665},
 primaryClass = {astro-ph.IM},
       adsurl = {https://ui.adsabs.harvard.edu/abs/2013PASP..125..306F}
}

@ARTICLE{Goodwin2022,
       author = {{Goodwin}, A.~J. and {van Velzen}, S. and {Miller-Jones}, J.~C.~A. and {Mummery}, A. and {Bietenholz}, M.~F. and {Wederfoort}, A. and {Hammerstein}, E. and {Bonnerot}, C. and {Hoffmann}, J. and {Yan}, L.},
        title = "{AT2019azh: an unusually long-lived, radio-bright thermal tidal disruption event}",
      journal = {\mnras},
         year = 2022,
        month = apr,
       volume = {511},
       number = {4},
        pages = {5328-5345},
          doi = {10.1093/mnras/stac333},
archivePrefix = {arXiv},
       eprint = {2201.03744},
 primaryClass = {astro-ph.HE},
       adsurl = {https://ui.adsabs.harvard.edu/abs/2022MNRAS.511.5328G}
}

@ARTICLE{Greene2020,
       author = {{Greene}, Jenny E. and {Strader}, Jay and {Ho}, Luis C.},
        title = "{Intermediate-Mass Black Holes}",
      journal = {\araa},
         year = 2020,
        month = aug,
       volume = {58},
        pages = {257-312},
          doi = {10.1146/annurev-astro-032620-021835},
archivePrefix = {arXiv},
       eprint = {1911.09678},
 primaryClass = {astro-ph.GA},
       adsurl = {https://ui.adsabs.harvard.edu/abs/2020ARA&A..58..257G}
}

@ARTICLE{Granot2002,
       author = {{Granot}, Jonathan and {Sari}, Re'em},
        title = "{The Shape of Spectral Breaks in Gamma-Ray Burst Afterglows}",
      journal = {\apj},
         year = 2002,
        month = apr,
       volume = {568},
       number = {2},
        pages = {820-829},
          doi = {10.1086/338966},
archivePrefix = {arXiv},
       eprint = {astro-ph/0108027},
 primaryClass = {astro-ph},
       adsurl = {https://ui.adsabs.harvard.edu/abs/2002ApJ...568..820G}
}

@ARTICLE{Greiner2015,
       author = {{Greiner}, Jochen and {Mazzali}, Paolo A. and {Kann}, D. Alexander and {Kr{\"u}hler}, Thomas and {Pian}, Elena and {Prentice}, Simon and {Olivares E.}, Felipe and {Rossi}, Andrea and {Klose}, Sylvio and {Taubenberger}, Stefan and {Knust}, Fabian and {Afonso}, Paulo M.~J. and {Ashall}, Chris and {Bolmer}, Jan and {Delvaux}, Corentin and {Diehl}, Roland and {Elliott}, Jonathan and {Filgas}, Robert and {Fynbo}, Johan P.~U. and {Graham}, John F. and {Guelbenzu}, Ana Nicuesa and {Kobayashi}, Shiho and {Leloudas}, Giorgos and {Savaglio}, Sandra and {Schady}, Patricia and {Schmidl}, Sebastian and {Schweyer}, Tassilo and {Sudilovsky}, Vladimir and {Tanga}, Mohit and {Updike}, Adria C. and {van Eerten}, Hendrik and {Varela}, Karla},
        title = "{A very luminous magnetar-powered supernova associated with an ultra-long {\ensuremath{\gamma}}-ray burst}",
      journal = {\nat},
         year = 2015,
        month = jul,
       volume = {523},
       number = {7559},
        pages = {189-192},
          doi = {10.1038/nature14579},
archivePrefix = {arXiv},
       eprint = {1509.03279},
 primaryClass = {astro-ph.HE},
       adsurl = {https://ui.adsabs.harvard.edu/abs/2015Natur.523..189G}
}

@ARTICLE{Guillochon2018,
       author = {{Guillochon}, James and {Nicholl}, Matt and {Villar}, V. Ashley and {Mockler}, Brenna and {Narayan}, Gautham and {Mandel}, Kaisey S. and {Berger}, Edo and {Williams}, Peter K.~G.},
        title = "{MOSFiT: Modular Open Source Fitter for Transients}",
      journal = {\apjs},
         year = 2018,
        month = may,
       volume = {236},
       number = {1},
          eid = {6},
        pages = {6},
          doi = {10.3847/1538-4365/aab761},
archivePrefix = {arXiv},
       eprint = {1710.02145},
 primaryClass = {astro-ph.IM},
       adsurl = {https://ui.adsabs.harvard.edu/abs/2018ApJS..236....6G}
}

@ARTICLE{Gupta2022,
       author = {{Gupta}, Rahul and {Kumar}, Amit and {Pandey}, Shashi Bhushan and {Castro-Tirado}, A.~J. and {Ghosh}, Ankur and {Dimple} and {Hu}, Y. -D. and {Fern{\'a}ndez-Garc{\'\i}a}, E. and {Caballero-Garc{\'\i}a}, M.~D. and {Castro-Tirado}, M. {\'A}. and {Hedrosa}, R.~P. and {Hermelo}, I. and {Vico}, I. and {Misra}, Kuntal and {Kumar}, Brajesh and {Aryan}, Amar and {Tiwari}, Sugriva Nath},
        title = "{Revealing nature of GRB 210205A, ZTF21aaeyldq (AT2021any) and follow-up observations with the 4K<inline-formula id=``IEq1''><mml:math><mml:mo>{\texttimes}</mml:mo></mml:math></inline-formula>4K CCD imager + 3.6m DOT}",
      journal = {Journal of Astrophysics and Astronomy},
         year = 2022,
        month = jun,
       volume = {43},
       number = {1},
          eid = {11},
        pages = {11},
          doi = {10.1007/s12036-021-09794-4},
archivePrefix = {arXiv},
       eprint = {2111.11795},
 primaryClass = {astro-ph.HE},
       adsurl = {https://ui.adsabs.harvard.edu/abs/2022JApA...43...11G}
}

@ARTICLE{Gutierrez2024,
       author = {{Guti{\'e}rrez}, Claudia P. and {Mattila}, Seppo and {Lundqvist}, Peter and {Dessart}, Luc and {Gonz{\'a}lez-Gait{\'a}n}, Santiago and {Jonker}, Peter G. and {Dong}, Subo and {Coppejans}, Deanne and {Chen}, Ping and {Charalampopoulos}, Panos and {Elias-Rosa}, Nancy and {Reynolds}, Thomas M. and {Kochanek}, Christopher and {Fraser}, Morgan and {Pastorello}, Andrea and {Gromadzki}, Mariusz and {Neustadt}, Jack and {Benetti}, Stefano and {Kankare}, Erkki and {Kangas}, Tuomas and {Kotak}, Rubina and {Stritzinger}, Maximilian D. and {Wevers}, Thomas and {Zhang}, Bing and {Bersier}, David and {Bose}, Subhash and {Buckley}, David A.~H. and {Dastidar}, Raya and {Gangopadhyay}, Anjasha and {Hamanowicz}, Aleksandra and {Kollmeier}, Juna A. and {Mao}, Jirong and {Misra}, Kuntal and {Potter}, Stephen. B. and {Prieto}, Jose L. and {Romero-Colmenero}, Encarni and {Singh}, Mridweeka and {Somero}, Auni and {Terreran}, Giacomo and {Vaisanen}, Petri and {Wyrzykowski}, {\L}ukasz},
        title = "{CSS 161010: A Luminous Fast Blue Optical Transient with Broad Blueshifted Hydrogen Lines}",
      journal = {\apj},
         year = 2024,
        month = dec,
       volume = {977},
       number = {2},
          eid = {162},
        pages = {162},
          doi = {10.3847/1538-4357/ad89a5},
archivePrefix = {arXiv},
       eprint = {2408.04698},
 primaryClass = {astro-ph.HE},
       adsurl = {https://ui.adsabs.harvard.edu/abs/2024ApJ...977..162G}
}

@ARTICLE{Hammerstein2026,
       author = {{Hammerstein}, Erica and {Cenko}, S. Bradley and {Andreoni}, Igor and {Charalampopoulos}, Panos and {Chornock}, Ryan and {Margutti}, Raffaella and {O'Connor}, Brendan and {Schulze}, Steve and {Sollerman}, Jesper and {Barway}, Sudhanshu and {Bhalerao}, Varun and {Anupama}, G.~C. and {Kumar}, Harsh and {Marini}, Ester and {Paris}, Diego and {Perley}, Daniel A. and {Rossi}, Andrea and {Yao}, Yuhan},
        title = "{The Jetted Tidal Disruption Event AT 2022cmc: Investigating Connections to the Optical Tidal Disruption Event Population and Spectral Subclasses through Late-time Follow-up}",
      journal = {\apj},
         year = 2026,
        month = jan,
       volume = {996},
       number = {2},
          eid = {143},
        pages = {143},
          doi = {10.3847/1538-4357/ae1838},
archivePrefix = {arXiv},
       eprint = {2506.08250},
 primaryClass = {astro-ph.HE},
       adsurl = {https://ui.adsabs.harvard.edu/abs/2026ApJ...996..143H}
}

@ARTICLE{Hatsukade2021APJL,
       author = {{Hatsukade}, B. and {Tominaga}, N. and {Morokuma}, T. and {Morokuma-Matsui}, K. and {Tamura}, Y. and {Niinuma}, K. and {Hayashi}, M. and {Matsuda}, Y. and {Motogi}, K.},
        title = "{Variability of Late-time Radio Emission in the Superluminous Supernova PTF10hgi}",
      journal = {\apjl},
         year = 2021,
        month = apr,
       volume = {911},
       number = {1},
          eid = {L1},
        pages = {L1},
          doi = {10.3847/2041-8213/abef03},
archivePrefix = {arXiv},
       eprint = {2103.09374},
 primaryClass = {astro-ph.HE},
       adsurl = {https://ui.adsabs.harvard.edu/abs/2021ApJ...911L...1H}
}

@ARTICLE{Ho2019,
       author = {{Ho}, Anna Y.~Q. and {Phinney}, E. Sterl and {Ravi}, Vikram and {Kulkarni}, S.~R. and {Petitpas}, Glen and {Emonts}, Bjorn and {Bhalerao}, V. and {Blundell}, Ray and {Cenko}, S. Bradley and {Dobie}, Dougal and {Howie}, Ryan and {Kamraj}, Nikita and {Kasliwal}, Mansi M. and {Murphy}, Tara and {Perley}, Daniel A. and {Sridharan}, T.~K. and {Yoon}, Ilsang},
        title = "{AT2018cow: A Luminous Millimeter Transient}",
      journal = {\apj},
         year = 2019,
        month = jan,
       volume = {871},
       number = {1},
          eid = {73},
        pages = {73},
          doi = {10.3847/1538-4357/aaf473},
archivePrefix = {arXiv},
       eprint = {1810.10880},
 primaryClass = {astro-ph.HE},
       adsurl = {https://ui.adsabs.harvard.edu/abs/2019ApJ...871...73H}
}

@ARTICLE{Ho2020,
       author = {{Ho}, Anna Y.~Q. and {Perley}, Daniel A. and {Kulkarni}, S.~R. and {Dong}, Dillon Z.~J. and {De}, Kishalay and {Chandra}, Poonam and {Andreoni}, Igor and {Bellm}, Eric C. and {Burdge}, Kevin B. and {Coughlin}, Michael and {Dekany}, Richard and {Feeney}, Michael and {Frederiks}, Dmitry D. and {Fremling}, Christoffer and {Golkhou}, V. Zach and {Graham}, Matthew J. and {Hale}, David and {Helou}, George and {Horesh}, Assaf and {Kasliwal}, Mansi M. and {Laher}, Russ R. and {Masci}, Frank J. and {Miller}, A.~A. and {Porter}, Michael and {Ridnaia}, Anna and {Rusholme}, Ben and {Shupe}, David L. and {Soumagnac}, Maayane T. and {Svinkin}, Dmitry S.},
        title = "{The Koala: A Fast Blue Optical Transient with Luminous Radio Emission from a Starburst Dwarf Galaxy at z = 0.27}",
      journal = {\apj},
         year = 2020,
        month = may,
       volume = {895},
       number = {1},
          eid = {49},
        pages = {49},
          doi = {10.3847/1538-4357/ab8bcf},
archivePrefix = {arXiv},
       eprint = {2003.01222},
 primaryClass = {astro-ph.HE},
       adsurl = {https://ui.adsabs.harvard.edu/abs/2020ApJ...895...49H}
}

@ARTICLE{Ho2022,
       author = {{Ho}, Anna Y.~Q. and {Margalit}, Ben and {Bremer}, Michael and {Perley}, Daniel A. and {Yao}, Yuhan and {Dobie}, Dougal and {Kaplan}, David L. and {O'Brien}, Andrew and {Petitpas}, Glen and {Zic}, Andrew},
        title = "{Luminous Millimeter, Radio, and X-Ray Emission from ZTF 20acigmel (AT 2020xnd)}",
      journal = {\apj},
         year = 2022,
        month = jun,
       volume = {932},
       number = {2},
          eid = {116},
        pages = {116},
          doi = {10.3847/1538-4357/ac4e97},
archivePrefix = {arXiv},
       eprint = {2110.05490},
 primaryClass = {astro-ph.HE},
       adsurl = {https://ui.adsabs.harvard.edu/abs/2022ApJ...932..116H}
}

@ARTICLE{Ho2023ApJ,
       author = {{Ho}, Anna Y.~Q. and {Perley}, Daniel A. and {Gal-Yam}, Avishay and {Lunnan}, Ragnhild and {Sollerman}, Jesper and {Schulze}, Steve and {Das}, Kaustav K. and {Dobie}, Dougal and {Yao}, Yuhan and {Fremling}, Christoffer and {Adams}, Scott and {Anand}, Shreya and {Andreoni}, Igor and {Bellm}, Eric C. and {Bruch}, Rachel J. and {Burdge}, Kevin B. and {Castro-Tirado}, Alberto J. and {Dahiwale}, Aishwarya and {De}, Kishalay and {Dekany}, Richard and {Drake}, Andrew J. and {Duev}, Dmitry A. and {Graham}, Matthew J. and {Helou}, George and {Kaplan}, David L. and {Karambelkar}, Viraj and {Kasliwal}, Mansi M. and {Kool}, Erik C. and {Kulkarni}, S.~R. and {Mahabal}, Ashish A. and {Medford}, Michael S. and {Miller}, A.~A. and {Nordin}, Jakob and {Ofek}, Eran and {Petitpas}, Glen and {Riddle}, Reed and {Sharma}, Yashvi and {Smith}, Roger and {Stewart}, Adam J. and {Taggart}, Kirsty and {Tartaglia}, Leonardo and {Tzanidakis}, Anastasios and {Winters}, Jan Martin},
        title = "{A Search for Extragalactic Fast Blue Optical Transients in ZTF and the Rate of AT2018cow-like Transients}",
      journal = {\apj},
         year = 2023,
        month = jun,
       volume = {949},
       number = {2},
          eid = {120},
        pages = {120},
          doi = {10.3847/1538-4357/acc533},
archivePrefix = {arXiv},
       eprint = {2105.08811},
 primaryClass = {astro-ph.HE},
       adsurl = {https://ui.adsabs.harvard.edu/abs/2023ApJ...949..120H}
}

@ARTICLE{Ho2023Nature,
       author = {{Ho}, Anna Y.~Q. and {Perley}, Daniel A. and {Chen}, Ping and {Schulze}, Steve and {Dhillon}, Vik and {Kumar}, Harsh and {Suresh}, Aswin and {Swain}, Vishwajeet and {Bremer}, Michael and {Smartt}, Stephen J. and {Anderson}, Joseph P. and {Anupama}, G.~C. and {Awiphan}, Supachai and {Barway}, Sudhanshu and {Bellm}, Eric C. and {Ben-Ami}, Sagi and {Bhalerao}, Varun and {de Boer}, Thomas and {Brink}, Thomas G. and {Burruss}, Rick and {Chandra}, Poonam and {Chen}, Ting-Wan and {Chen}, Wen-Ping and {Cooke}, Jeff and {Coughlin}, Michael W. and {Das}, Kaustav K. and {Drake}, Andrew J. and {Filippenko}, Alexei V. and {Freeburn}, James and {Fremling}, Christoffer and {Fulton}, Michael D. and {Gal-Yam}, Avishay and {Galbany}, Llu{\'\i}s and {Gao}, Hua and {Graham}, Matthew J. and {Gromadzki}, Mariusz and {Guti{\'e}rrez}, Claudia P. and {Hinds}, K. -Ryan and {Inserra}, Cosimo and {A J}, Nayana and {Karambelkar}, Viraj and {Kasliwal}, Mansi M. and {Kulkarni}, Shri and {M{\"u}ller-Bravo}, Tom{\'a}s E. and {Magnier}, Eugene A. and {Mahabal}, Ashish A. and {Moore}, Thomas and {Ngeow}, Chow-Choong and {Nicholl}, Matt and {Ofek}, Eran O. and {Omand}, Conor M.~B. and {Onori}, Francesca and {Pan}, Yen-Chen and {Pessi}, Priscila J. and {Petitpas}, Glen and {Polishook}, David and {Poshyachinda}, Saran and {Pursiainen}, Miika and {Riddle}, Reed and {Rodriguez}, Antonio C. and {Rusholme}, Ben and {Segre}, Enrico and {Sharma}, Yashvi and {Smith}, Ken W. and {Sollerman}, Jesper and {Srivastav}, Shubham and {Strotjohann}, Nora Linn and {Suhr}, Mark and {Svinkin}, Dmitry and {Wang}, Yanan and {Wiseman}, Philip and {Wold}, Avery and {Yang}, Sheng and {Yang}, Yi and {Yao}, Yuhan and {Young}, David R. and {Zheng}, WeiKang},
        title = "{Minutes-duration optical flares with supernova luminosities}",
      journal = {\nat},
         year = 2023,
        month = nov,
       volume = {623},
       number = {7989},
        pages = {927-931},
          doi = {10.1038/s41586-023-06673-6},
archivePrefix = {arXiv},
       eprint = {2311.10195},
 primaryClass = {astro-ph.HE},
       adsurl = {https://ui.adsabs.harvard.edu/abs/2023Natur.623..927H}
}

@ARTICLE{Ibik2024,
       author = {{Ibik}, Adaeze L. and {Drout}, Maria R. and {Margutti}, Raffaela and {Matthews}, David and {Villar}, V. Ashley and {Berger}, Edo and {Chornock}, Ryan and {Alexander}, Kate D. and {Eftekhari}, Tarraneh and {Laskar}, Tanmoy and {Lunnan}, Ragnhild and {Foley}, Ryan J. and {Jones}, David and {Milisavljevic}, Dan and {Rest}, Armin and {Scolnic}, Daniel and {Williams}, Peter K.~G.},
        title = "{PS1-11aop: Probing the Mass Loss History of a Luminous Interacting Supernova Prior to its Final Eruption with Multi-wavelength Observations}",
      journal = {arXiv e-prints},
         year = 2024,
        month = oct,
          eid = {arXiv:2410.15140},
        pages = {arXiv:2410.15140},
          doi = {10.48550/arXiv.2410.15140},
archivePrefix = {arXiv},
       eprint = {2410.15140},
 primaryClass = {astro-ph.HE},
       adsurl = {https://ui.adsabs.harvard.edu/abs/2024arXiv241015140I}
}

@ARTICLE{Inserra2018,
       author = {{Inserra}, C. and {Prajs}, S. and {Gutierrez}, C.~P. and {Angus}, C. and {Smith}, M. and {Sullivan}, M.},
        title = "{A Statistical Approach to Identify Superluminous Supernovae and Probe Their Diversity}",
      journal = {\apj},
         year = 2018,
        month = feb,
       volume = {854},
       number = {2},
          eid = {175},
        pages = {175},
          doi = {10.3847/1538-4357/aaaaaa},
archivePrefix = {arXiv},
       eprint = {1711.03787},
 primaryClass = {astro-ph.HE},
       adsurl = {https://ui.adsabs.harvard.edu/abs/2018ApJ...854..175I}
}

@ARTICLE{Jin2025,
       author = {{Jin}, C. -C. and {Li}, D. -Y. and {Jiang}, N. and {Dai}, L. -X. and {Cheng}, H. -Q. and {Zhu}, J. -Z. and {Yang}, C. -W. and {Rau}, A. and {Baldini}, P. and {Wang}, T. -G. and et al.},
        title = "{An Intermediate-mass Black Hole Lurking in A Galactic Halo Caught Alive during Outburst}",
      journal = {arXiv e-prints},
         year = 2025,
        month = jan,
          eid = {arXiv:2501.09580},
        pages = {arXiv:2501.09580},
          doi = {10.48550/arXiv.2501.09580},
archivePrefix = {arXiv},
       eprint = {2501.09580},
 primaryClass = {astro-ph.HE},
       adsurl = {https://ui.adsabs.harvard.edu/abs/2025arXiv250109580J}
}

@ARTICLE{Kann2011,
       author = {{Kann}, D.~A. and {Klose}, S. and {Zhang}, B. and {Covino}, S. and {Butler}, N.~R. and {Malesani}, D. and {Nakar}, E. and {Wilson}, A.~C. and {Antonelli}, L.~A. and {Chincarini}, G. and {Cobb}, B.~E. and {D'Avanzo}, P. and {D'Elia}, V. and {Della Valle}, M. and {Ferrero}, P. and {Fugazza}, D. and {Gorosabel}, J. and {Israel}, G.~L. and {Mannucci}, F. and {Piranomonte}, S. and {Schulze}, S. and {Stella}, L. and {Tagliaferri}, G. and {Wiersema}, K.},
        title = "{The Afterglows of Swift-era Gamma-Ray Bursts. II. Type I GRB versus Type II GRB Optical Afterglows}",
      journal = {\apj},
         year = 2011,
        month = jun,
       volume = {734},
       number = {2},
          eid = {96},
        pages = {96},
          doi = {10.1088/0004-637X/734/2/96},
archivePrefix = {arXiv},
       eprint = {0804.1959},
 primaryClass = {astro-ph},
       adsurl = {https://ui.adsabs.harvard.edu/abs/2011ApJ...734...96K}
}

@ARTICLE{Kare2021,
       author = {{Kale}, Ruta and {Ishwara-Chandra}, C.~H.},
        title = "{CAPTURE: a continuum imaging pipeline for the uGMRT}",
      journal = {Experimental Astronomy},
         year = 2021,
        month = feb,
       volume = {51},
       number = {1},
        pages = {95-108},
          doi = {10.1007/s10686-020-09677-6},
archivePrefix = {arXiv},
       eprint = {2010.00196},
 primaryClass = {astro-ph.IM},
       adsurl = {https://ui.adsabs.harvard.edu/abs/2021ExA....51...95K}
}

@ARTICLE{Klencki2025,
       author = {{Klencki}, Jakub and {Metzger}, Brian D.},
        title = "{Luminous Fast Blue Optical Transients as ``Failed'' Gravitational Wave Sources: Helium Core$-$Black Hole Mergers Following Delayed Dynamical Instability}",
      journal = {arXiv e-prints},
         year = 2025,
        month = oct,
          eid = {arXiv:2510.09745},
        pages = {arXiv:2510.09745},
          doi = {10.48550/arXiv.2510.09745},
archivePrefix = {arXiv},
       eprint = {2510.09745},
 primaryClass = {astro-ph.HE},
       adsurl = {https://ui.adsabs.harvard.edu/abs/2025arXiv251009745K}
}

@ARTICLE{Konig2022,
       author = {{K{\"o}nig}, O. and {Saxton}, R.~D. and {Kretschmar}, P. and {Angelini}, L. and {Belanger}, G. and {Evans}, P.~A. and {Freyberg}, M.~J. and {Savchenko}, V. and {Traulsen}, I. and {Wilms}, J.},
        title = "{HILIGT, Upper Limit Servers II - Implementing the data servers}",
      journal = {Astronomy and Computing},
         year = 2022,
        month = jan,
       volume = {38},
          eid = {100529},
        pages = {100529},
          doi = {10.1016/j.ascom.2021.100529},
archivePrefix = {arXiv},
       eprint = {2111.13563},
 primaryClass = {astro-ph.HE},
       adsurl = {https://ui.adsabs.harvard.edu/abs/2022A&C....3800529K}
}

@ARTICLE{Kormendy2013,
       author = {{Kormendy}, John and {Ho}, Luis C.},
        title = "{Coevolution (Or Not) of Supermassive Black Holes and Host Galaxies}",
      journal = {\araa},
         year = 2013,
        month = aug,
       volume = {51},
       number = {1},
        pages = {511-653},
          doi = {10.1146/annurev-astro-082708-101811},
archivePrefix = {arXiv},
       eprint = {1304.7762},
 primaryClass = {astro-ph.CO},
       adsurl = {https://ui.adsabs.harvard.edu/abs/2013ARA&A..51..511K}
}

@ARTICLE{Kriek2009,
       author = {{Kriek}, Mariska and {van Dokkum}, Pieter G. and {Labb{\'e}}, Ivo and {Franx}, Marijn and {Illingworth}, Garth D. and {Marchesini}, Danilo and {Quadri}, Ryan F.},
        title = "{An Ultra-Deep Near-Infrared Spectrum of a Compact Quiescent Galaxy at z = 2.2}",
      journal = {\apj},
         year = 2009,
        month = jul,
       volume = {700},
       number = {1},
        pages = {221-231},
          doi = {10.1088/0004-637X/700/1/221},
archivePrefix = {arXiv},
       eprint = {0905.1692},
 primaryClass = {astro-ph.CO},
       adsurl = {https://ui.adsabs.harvard.edu/abs/2009ApJ...700..221K}
}

@ARTICLE{Kuin2019,
       author = {{Kuin}, N. Paul M. and {Wu}, Kinwah and {Oates}, Samantha and {Lien}, Amy and {Emery}, Sam and {Kennea}, Jamie A. and {de Pasquale}, Massimiliano and {Han}, Qin and {Brown}, Peter J. and {Tohuvavohu}, Aaron and {Breeveld}, Alice and {Burrows}, David N. and {Cenko}, S. Bradley and {Campana}, Sergio and {Levan}, Andrew and {Markwardt}, Craig and {Osborne}, Julian P. and {Page}, Mat J. and {Page}, Kim L. and {Sbarufatti}, Boris and {Siegel}, Michael and {Troja}, Eleonora},
        title = "{Swift spectra of AT2018cow: a white dwarf tidal disruption event?}",
      journal = {\mnras},
         year = 2019,
        month = aug,
       volume = {487},
       number = {2},
        pages = {2505-2521},
          doi = {10.1093/mnras/stz053},
archivePrefix = {arXiv},
       eprint = {1808.08492},
 primaryClass = {astro-ph.HE},
       adsurl = {https://ui.adsabs.harvard.edu/abs/2019MNRAS.487.2505K}
}

@ARTICLE{Lacy2020,
       author = {{Lacy}, M. and {Baum}, S.~A. and {Chandler}, C.~J. and {Chatterjee}, S. and {Clarke}, T.~E. and {Deustua}, S. and {English}, J. and {Farnes}, J. and {Gaensler}, B.~M. and {Gugliucci}, N. and {Hallinan}, G. and {Kent}, B.~R. and {Kimball}, A. and {Law}, C.~J. and {Lazio}, T.~J.~W. and {Marvil}, J. and {Mao}, S.~A. and {Medlin}, D. and {Mooley}, K. and {Murphy}, E.~J. and {Myers}, S. and {Osten}, R. and {Richards}, G.~T. and {Rosolowsky}, E. and {Rudnick}, L. and {Schinzel}, F. and {Sivakoff}, G.~R. and {Sjouwerman}, L.~O. and {Taylor}, R. and {White}, R.~L. and {Wrobel}, J. and {Andernach}, H. and {Beasley}, A.~J. and {Berger}, E. and {Bhatnager}, S. and {Birkinshaw}, M. and {Bower}, G.~C. and {Brandt}, W.~N. and {Brown}, S. and {Burke-Spolaor}, S. and {Butler}, B.~J. and {Comerford}, J. and {Demorest}, P.~B. and {Fu}, H. and {Giacintucci}, S. and {Golap}, K. and {G{\"u}th}, T. and {Hales}, C.~A. and {Hiriart}, R. and {Hodge}, J. and {Horesh}, A. and {Ivezi{\'c}}, {\v{Z}}. and {Jarvis}, M.~J. and {Kamble}, A. and {Kassim}, N. and {Liu}, X. and {Loinard}, L. and {Lyons}, D.~K. and {Masters}, J. and {Mezcua}, M. and {Moellenbrock}, G.~A. and {Mroczkowski}, T. and {Nyland}, K. and {O'Dea}, C.~P. and {O'Sullivan}, S.~P. and {Peters}, W.~M. and {Radford}, K. and {Rao}, U. and {Robnett}, J. and {Salcido}, J. and {Shen}, Y. and {Sobotka}, A. and {Witz}, S. and {Vaccari}, M. and {van Weeren}, R.~J. and {Vargas}, A. and {Williams}, P.~K.~G. and {Yoon}, I.},
        title = "{The Karl G. Jansky Very Large Array Sky Survey (VLASS). Science Case and Survey Design}",
      journal = {\pasp},
         year = 2020,
        month = mar,
       volume = {132},
       number = {1009},
          eid = {035001},
        pages = {035001},
          doi = {10.1088/1538-3873/ab63eb},
archivePrefix = {arXiv},
       eprint = {1907.01981},
 primaryClass = {astro-ph.IM},
       adsurl = {https://ui.adsabs.harvard.edu/abs/2020PASP..132c5001L}
}

@ARTICLE{Levine2023,
       author = {{Levine}, D. and {Dainotti}, M. and {Fraija}, N. and {Warren}, D. and {Chandra}, P. and {Lloyd-Ronning}, N.},
        title = "{Interpretation of radio afterglows in the framework of the standard fireball and energy injection models}",
      journal = {\mnras},
         year = 2023,
        month = mar,
       volume = {519},
       number = {3},
        pages = {4670-4683},
          doi = {10.1093/mnras/stac3730},
archivePrefix = {arXiv},
       eprint = {2212.07971},
 primaryClass = {astro-ph.HE},
       adsurl = {https://ui.adsabs.harvard.edu/abs/2023MNRAS.519.4670L}
}

@ARTICLE{Li2025AT2023lcr,
       author = {{Li}, Maggie L. and {Ho}, Anna Y.~Q. and {Ryan}, Geoffrey and {Perley}, Daniel A. and {Lamb}, Gavin P. and {Nayana}, A.~J. and {Andreoni}, Igor and {Anupama}, G.~C. and {Bellm}, Eric C. and {Berger}, Edo and {Bloom}, Joshua S. and {Burns}, Eric and {Caiazzo}, Ilaria and {Chandra}, Poonam and {Coughlin}, Michael W. and {El-Badry}, Kareem and {Graham}, Matthew J. and {Kasliwal}, Mansi and {Keating}, Garrett K. and {Kulkarni}, S.~R. and {Kumar}, Harsh and {Masci}, Frank J. and {Perley}, Richard A. and {Purdum}, Josiah and {Rao}, Ramprasad and {Rodriguez}, Antonio C. and {Rusholme}, Ben and {Sarin}, Nikhil and {Sollerman}, Jesper and {Srinivasaragavan}, Gokul P. and {Swain}, Vishwajeet and {Vanderbosch}, Zachary},
        title = "{The Nature of Optical Afterglows without Gamma-Ray Bursts: Identification of AT2023lcr and Multiwavelength Modeling}",
      journal = {\apj},
         year = 2025,
        month = may,
       volume = {985},
       number = {1},
          eid = {124},
        pages = {124},
          doi = {10.3847/1538-4357/adc800},
archivePrefix = {arXiv},
       eprint = {2411.07973},
 primaryClass = {astro-ph.HE},
       adsurl = {https://ui.adsabs.harvard.edu/abs/2025ApJ...985..124L}
}

@ARTICLE{Li2025,
       author = {{Li}, D.-Y. and {Zhang}, W.-D. and {Yang}, J. and {Chen}, J.-H. and {Yuan}, W. and {Cheng}, H.-Q. and {Xu}, F. and {Shu}, X.-W. and {Shen}, R.-F. and {Jiang}, N. and {Zhu}, J.-Z. and {Zhou}, C. and {Lei}, W.-H. and {Sun}, H. and {Jin}, C.-C. and {Dai}, L.-X. and {Zhang}, B. and {Yang}, Y.-H. and {Zhang}, W.-J. and {Feng}, H. and {Liu}, B.-F. and {Zhou}, H.-Y. and {Pan}, H.-W. and {Liu}, M.-J. and {Corbel}, S. and {Jagan}, S.~K. and {Baglio}, M.~C. and {Burns}, C. and {Cangemi}, F. and {Chen}, C. and {Cheng}, Y.-H. and {Coleiro}, A. and {Coti Zelati}, F. and {Das}, S.~R. and {Dong}, Z.-N. and {Galbany}, L. and {Grollimund}, N. and {Kelson}, D. and {Lai}, D. and {Li}, X. and {Liu}, Y. and {Marino}, A. and {Mockler}, B. and {O'Brien}, P. and {Qiao}, E.-L. and {Rea}, N. and {Resmi}, L. and {Rodriguez}, J. and {Saxton}, R. and {Sun}, L.-M. and {Tao}, L. and {Wang}, T.-G. and {Wang}, Y.-L. and {Wu}, X.-F. and {Xu}, D. and {Zhang}, Y.-J. and {Zhao}, G.-Y. and {Cai}, Z.-M. and {Chen}, Y. and {Cordier}, B. and {Fan}, Z. and {Gao}, H. and {Ghirlanda}, G. and {Hu}, J.-W. and {Huang}, Y.-F. and {Jia}, S.-M. and {Komossa}, S. and {Liu}, H.-Y. and {Liu}, H.-Q. and {Ness}, J.-U. and {Rau}, A. and {Sanders}, J. and {Soria}, R. and {Sun}, S.-L. and {Sun}, X.-J. and {Troja}, E. and {Wen}, S.-X. and {Xue}, Y.-Q. and {Yin}, Y.-H.~I. and {Zhang}, C. and {Zhang}, S.-N. and {Zhang}, Y.-H.},
        title = "{A fast powerful X-ray transient from possible tidal disruption of a white dwarf}",
      journal = {arXiv e-prints},
         year = 2025,
        month = sep,
          eid = {arXiv:2509.25877},
        pages = {arXiv:2509.25877},
          doi = {10.48550/arXiv.2509.25877},
archivePrefix = {arXiv},
       eprint = {2509.25877},
 primaryClass = {astro-ph.HE},
       adsurl = {https://ui.adsabs.harvard.edu/abs/2025arXiv250925877L}
}

@ARTICLE{Lin2018,
       author = {{Lin}, Dacheng and {Strader}, Jay and {Carrasco}, Eleazar R. and {Page}, Dany and {Romanowsky}, Aaron J. and {Homan}, Jeroen and {Irwin}, Jimmy A. and {Remillard}, Ronald A. and {Godet}, Olivier and {Webb}, Natalie A. and {Baumgardt}, Holger and {Wijnands}, Rudy and {Barret}, Didier and {Duc}, Pierre-Alain and {Brodie}, Jean P. and {Gwyn}, Stephen D.~J.},
        title = "{A luminous X-ray outburst from an intermediate-mass black hole in an off-centre star cluster}",
      journal = {Nature Astronomy},
         year = 2018,
        month = jun,
       volume = {2},
        pages = {656-661},
          doi = {10.1038/s41550-018-0493-1},
archivePrefix = {arXiv},
       eprint = {1806.05692},
 primaryClass = {astro-ph.HE},
       adsurl = {https://ui.adsabs.harvard.edu/abs/2018NatAs...2..656L}
}

@ARTICLE{Nordin2019TNS,
       author = {{Nordin}, J. and {Brinnel}, V. and {Giomi}, M. and {Santen}, J.~V. and {Gal-Yam}, A. and {Yaron}, O. and {Schulze}, S.},
        title = "{ZTF Transient Discovery Report for 2019-06-28}",
      journal = {Transient Name Server Discovery Report},
         year = 2019,
        month = jun,
       volume = {2019-1095},
        pages = {1},
       adsurl = {https://ui.adsabs.harvard.edu/abs/2019TNSTR1095....1N}
}

\end{document}